\documentclass[aps, prd, preprintnumbers, floatfix, superscriptaddress, 10pt]{revtex4}

\usepackage[dvips]{graphics}
\usepackage{amsmath,amsfonts,bm}
\usepackage{epsfig,bm}
\usepackage{graphicx,comment}
\usepackage{dcolumn,color}

\usepackage{latexsym}
\usepackage{colordvi}
\usepackage{amssymb}

 \def\cN{{\cal N}} \def\cO{{\cal O}} 
   
 \def\cV{{\cal V}}  
 
   \def\GeV{{\rm GeV}}       
                
\def\sla{\slash{\!\!\!}}

\def\lsim{\mathrel{\rlap{\lower3pt\hbox{\hskip1pt$\sim$}}
     \raise1pt\hbox{$<$}}} 
\def\gsim{\mathrel{\rlap{\lower3pt\hbox{\hskip1pt$\sim$}}
     \raise1pt\hbox{$>$}}}

\def\be{\begin{eqnarray}}\def\ba{\begin{eqnarray}}
\def\ee{\end{eqnarray}}\def\ea{\end{eqnarray}}
\def\ben{\begin{enumerate}}\def\bitem{\begin{itemize}}
\def\een{\end{enumerate}}\def\eitem{\end{itemize}}

\begin{document}



\preprint{\parbox[b]{1in}{ \hbox{\tt IC/2009/012} }}
\title{Electromagnetic Nucleon-to-Delta Transition in Holographic QCD}


\author{Hovhannes R. Grigoryan}
\email[E-mail: ]{grigoryan@phy.anl.gov} 
\affiliation{Physics Division, Argonne National Laboratory, Argonne, Illinois 60439, USA} \vspace{0.1in}

\author{T.-S.H. Lee }
\email[E-mail: ]{lee@phy.anl.gov} 
 \affiliation{Physics Division, Argonne National Laboratory, Argonne, Illinois 60439, USA} \vspace{0.1in}

\author{Ho-Ung Yee}
\email[E-mail: ]{hyee@ictp.it} 
 \affiliation{ICTP, High Energy, Cosmology and Astroparticle Physics, Strada Costiera 11, 34014, Trieste, Italy}

\begin{abstract}
We study nucleon-to-delta electromagnetic transition form factors and relations between them within
the framework of the holographic dual model of QCD proposed by Sakai and Sugimoto.
In this setup, baryons appear as topological solitons of the five-dimensional holographic
gauge theory that describes a tower of mesons and their interactions.
We find a relativistic extension of the nucleon-delta-vector meson interaction vertices and use these to calculate transition
form factors from holographic QCD.
We observe that at low momentum transfer, magnetic dipole, electric and Coulomb quadrupole form factors
and their ratios follow the patterns expected in the large $N_c$ limit.
Our results at this approximation are in reasonable agreement with experiment.
\end{abstract}


\keywords{QCD, AdS-CFT Correspondence}
\pacs{11.25.Tq, 
11.10.Kk, 
11.15.Tk  
12.38.Lg  
}

\maketitle


\section{Introduction}

As is well known, $\Delta(1232)$ resonance is the first excited state of the nucleon and plays an important role in
strong interaction physics. Experimentally, $\Delta$'s are produced in scattering pions or electron beams off a nucleon target.
The $\Delta(1232)$ has isospin 3/2 and therefore comes in four different charge states: $\Delta^{++}$, $\Delta^+$, $\Delta^0$, and
$\Delta^-$ with approximately the same mass and width.
The spin of the $\Delta(1232)$ is also 3/2, and it is the lightest known particle with such a spin.
It appears that it decays via $\Delta \to N\pi$ with $99\%$ branching ratio \cite{Amsler:2008zzb}, and only less than $1\%$ to
the total decay width is coming from the EM channel ($\Delta \to N\gamma$).
This EM $\gamma N \Delta$ transition is predominantly of the magnetic dipole ($M1$) type.


With the lack of complete theoretical control over nonperturbative low energy QCD, at present, one should explore various complementary
techniques to gain more understanding on nonperturbative QCD phenomena. The large $N_c$ limit is one such attempt that has been
pursued for some time, and it was revived a decade ago by a proposal that in conjunction with the idea of holography, the large $N_c$ QCD
in the strong coupling regime may be described by a weakly coupled dual five-dimensional (5D) model, with the additional fifth direction playing the role of the energy scale \cite{Maldacena:1997re}. Although the precise dual model of large $N_c$ QCD has not been found, several approximate models
were proposed applying both so-called ``top-down'' \cite{Sakai:2004cn} and ``bottom-up'' \cite{Polchinski:2002jw} approaches.
The study of hadronic form factors in these models of holographic QCD
\cite{Sakai:2004cn}--\cite{Hong:2007kx}
allowed us to gain a confidence that the holographic QCD approach can be a useful complementary tool in predicting the low energy
behavior of QCD at least in the large $N_c$ limit.

In this work, our goal is to go one step further and investigate the $\gamma^*N \to \Delta$ transition form factors in the framework of
holographic QCD. We will work in the top-down model proposed by Sakai and Sugimoto \cite{Sakai:2004cn}.
Baryons in this model have been studied recently \cite{Nawa:2006gv}--\cite{Kim:2008iy} and our methods are based on some of
these developments.
In particular, we start our analysis by considering the nonrelativistic nucleon-delta-vector meson vertices found by Park-Yi \cite{Park:2008sp}, and for completeness also discuss the related approach by Hashimoto-Sakai-Sugimoto \cite{Hashimoto:2008zw}.

Using the nonrelativistic result in Ref.~\cite{Park:2008sp}, we find a relativistic generalization of the nucleon-delta-vector meson vertex which is required for consistent treatment of transition form factors.
This in turn is essential for comparison of the model predictions for nucleon-delta transition form factors with experiment.
The knowledge of these form factors is proven to be an important and complex check for any model of strong interactions.
Thus, we would like to investigate what holographic QCD can tell us about this process and how well
its predictions agree with experiment.

Since the virtual photon has three polarizations, the $\gamma^*N \to \Delta$ transition should be in general described by three
independent form factors. These three form factors are related to the magnetic dipole ($M1$), electric ($E2$) and Coulomb ($C2$)
quadrupole types of transitions.
The $\gamma N\Delta$ transition was measured in the pion photoproduction and electroproduction reactions in the $\Delta$-resonance energy
region. The $E2$ and $C2$ transitions were found to be relatively small but nonzero at moderate  momentum transfers $Q^2$, with the
ratios $R_{EM} = E2/M1$ and $R_{SM}=C2/M1 $ being at the level of a few percent.
The smallness of these ratios seems to have a purely nonperturbative origin. Indeed, the perturbative QCD studies \cite{Carlson:1985mm}
predict the same strength for $E2$ and $M1$ transitions at asymptotically large $Q^2$, while experimentally, the $E2/M1$ ratio
is negative and very close to zero for energies up to $ Q^2 \sim 4 \
\GeV^2$ \cite{Beck:1997ew}.

The smallness of the $E2/M1$ ratio is a very well-known prediction of the quark model, where the $N\to \Delta$ transition is described
by a spin flip of a quark in the $s$-wave state, which in the $\gamma N \Delta$ case leads to the $M1$ type of transition.
Any $d$-wave admixture in the wave function of $\Delta$ would allow for the $E2$ and $C2$ quadrupole transitions. Therefore, by
measuring these transitions, one is able to observe the presence of the $d$-wave components and hence quantify to which extent the
nucleon or the $\Delta$ wave function deviates from the spherical shape.

Within the nonrelativistic SU(6) quark model it was shown that $E2$ is zero \cite{Becchi:1965} provided that quarks have zero orbital
angular momentum. Small values for $E2/M1$ in the region $Q^2 < 4 \
\GeV^2$ were obtained in the relativistic quark model, see e.g. Ref.~\cite{Aznaurian:1993rk}.
In the large $N_c$ limit of QCD, it was shown that the $E2/M1$ ratio is of order $\cO(1/N_c^2)$ \cite{Jenkins:2002rj} without any
assumption about the quark orbital angular momentum or intrinsic deformation of the baryon. We will show below that the same features can be also reproduced from the holographic QCD.

There have been studies on the $\gamma^*N \to \Delta$ transition form factors using other methods, for example, the local quark-hadron
duality approach motivated by QCD sum rules \cite{Belyaev:1995ya} and the framework of the light-cone sum rules in
\cite{Braun:2005be}. Another approach using the method of QCD sum rules is given in Ref.~\cite{Wang:2009ru}.
Recent lattice calculations \cite{Alexandrou:2004xn} of the $N
\Delta$ transition form factors up to $1.5 \ \GeV^2$ give small negative values for the ratio $E2/M1$. All these results provide
strong evidence that the observed small  value of $E2/M1$ has a purely nonperturbative origin.
This is why we think that the application of the holographic QCD may shed more light on our understanding of this phenomenon, as it
captures both nonperturbative and large $N_c$ features.
Some of the reviews describing various aspects of the $\gamma N \Delta$ transition can be found in \cite{Review}.


The paper is organized as follows. In Sec.~2, we briefly review the holographic model proposed by Sakai and Sugimoto \cite{Sakai:2004cn}, and discuss how baryons are described in the model. Instead of going into the details of this construction that are well described in
Refs.~\cite{Sakai:2004cn,Sakai:2005yt}, we only outline the five-dimensional effective action that emerges from the model and
the field decomposition which are of importance for our further discussions.
We then review the primary results in Ref.~\cite{Park:2008sp}, where nucleons and delta baryons as well as their interactions with the
vector mesons are studied within the nonrelativistic formalism. This will be our starting point of the subsequent analysis.
In Sec.~3, we explore the holographic vector meson dominance feature that emerges from the holographic QCD. In the origin of this
lies the observation that there exists a basis for the vector meson fields in which the external electromagnetic field interacts only
with vector meson fields, linearly without kinetic mixing. As a result, we end up with a very convenient framework to perform
tree-level calculations.
In Sec.~4, we outline similar approaches for describing baryons in the Sakai-Sugimoto model (other than Ref.~\cite{Park:2008sp}).
For completeness, we also study and compare the dependence of some of the main results,
when applying these alternative approaches.

In Sec.~5, we propose a relativistic generalization for the $N\Delta
v^{(n)}$ vertex that is required to obtain relativistic transition form factors. Since our purpose is to perform tree level
calculations, we avoid all problems associated with the higher-spin fermions. Without concerning what is the appropriate
5D relativistic formulation for the spin-$3/2$ fermions, we simply explore the expectation that after integrating
over the holographic 5'th direction $Z$, the theory should effectively become a 4D relativistic theory of
spin-$3/2$ particles, for which Rarita-Schwinger formalism can be implemented. This in turn should reduce to a nonrelativistic theory
with $N\Delta v^{(n)}$ vertex that we started with.
By considering all possible relativistic operators consistent with 4D symmetries, we eventually find a unique relativistic
operator which satisfies these requirements, and write down a 4D relativistic Lagrangian that correctly describes the
$N\Delta v^{(n)}$ vertices, up to possible subleading terms of order $\cO(1/N_c)$.

In Sec.~6, after giving a general formalism for the interaction vertex and defining the relevant form factors, together with their
ratios that are of importance in comparison with experiments, we present our results for the $\gamma N\Delta$ transition form factors
from the holographic QCD. We observe that magnetic dipole, electric, and Coulomb quadrupole form factors all depend on a single
\emph{holographic form factor}.
Similar observation was also made in Ref.\cite{Hashimoto:2008zw}.
This leads to interesting consequences that are in accord with the
expected behavior in the large $N_c$ limit. We also briefly discuss some
other observables of interest.
We conclude by summarizing our results and pointing out some directions for further development.


\section{Preliminaries}

We briefly outline the holographic model proposed by Sakai and
Sugimoto in Ref.~\cite{Sakai:2004cn} in order to establish our
notations and conventions that we are going to use throughout this
paper. Although we give a very short description of the original construction,
we stress that practical readers can simply start from the resulting
5D gauge theory given in (\ref{5dtheory}) and
(\ref{CS5d}), without referring to the details of the construction. We
will also make a few comments about the holographic treatment of
baryons and conclude by describing the nonrelativistic
nucleon-$\Delta$-vector meson vertex obtained originally in
Ref.~\cite{Park:2008sp}.

\subsection{Model of Sakai and Sugimoto}

The Sakai-Sugimoto model is constructed by placing probe $N_f$ $D8$ and $\overline{D8}$-branes in the $S^1$ compactified $N_c$
$D4$-brane background of Type IIA string theory.
In the $D$-brane picture, the supersymmetry on a $D4$-brane is broken by imposing an antiperiodic
boundary conditions on fermions along the $S^1$ circle.
As a result, for energies lower than the compactification scale, we are effectively left with the 4D pure $SU(N_c)$
gauge theory.
In accordance with the basics of AdS/CFT \cite{Maldacena:1997re}, in the large $N_c$ limit, the theory describing $N_c$ coincident $D4$-branes is expected to be dual to a supergravity theory in a curved background with flux obtained by solving Type IIA supergravity.

Working in the probe approximation, that is when $N_f/N_c \ll 1 $,
one can ignore the backreaction of $D8$ and $\overline{D8}$ branes
embedded in the $D4$-brane background. The introduction of $D8$- and
$\overline{D8}$-branes supports the existence of $N_f$ massless flavors of quarks.
In this model, quarks of left- and right-handed chiralities appear from $D8-D4$ and $\overline{D8}-D4$ strings respectively.
This construction aims to reproduce large $N_c$ QCD with $N_f$ massless quarks.
The chiral $U(N_f)_L \times U(N_f)_R$ symmetry group of QCD emerges
from the $U(N_f)_L \times U(N_f)_R$ gauge theory living on the probe
$D8$- and $\overline{D8}$-branes. One expects that the dynamics of
this gauge theory, living on the $D8$ and $\overline{D8}$ branes,
holographically describes the chiral dynamics of the large $N_c$ QCD
with massless quarks.

Geometrically, the chiral $U(N_f)_L \times U(N_f)_R$ symmetry of QCD
is spontaneously broken to the diagonal subgroup $U(N_f)_V$ due to
the ``merging'' of these $D8$ and $\overline{D8}$ branes in the
background of $D4$-branes. In simple terms, the resulting configuration can be
viewed as a stack of $N_f$ $D8$-branes continuously connecting asymptotic
regions of the original $D8$ and $\overline{D8}$ branes. The
Nambu-Goldstone bosons of the broken chiral symmetry arise from the
Wilson line that connects these two asymptotically separated regions
on the $D8$-brane. Other modes on the $D8$-branes correspond to a
tower of vector and axial-vector mesons, whose interactions among
themselves are completely determined from the theory describing $D8$-branes.


The resulting 5D theory on the $D8$-branes is a $U(N_f)$ non-Abelian gauge theory with a Chern-Simons (CS) term in a curved background metric. In addition to the usual 4D Minkowski spacetime coordinates, there is the holographic dimension $Z$, that ranges from $-\infty$ to $+\infty$. The global $U(N_f)_L$ and $U(N_f)_R$ chiral symmetries of QCD reside at the boundaries $Z=+\infty$ and $Z=-\infty$, respectively.
More precisely, gauge fields on these boundaries only couple to the left or right chiral currents of QCD.
The action of this 5D gauge theory consists of a gauge-kinetic term, that can be written as
\be
S_{D8}={\pi\over 4}{\left(f_\pi\over M_{KK}\right)^2} \int d^4
x\, dZ\,\, {\rm Tr}\left(-{1\over 2} \left(1+Z^2\right)^{-{1\over
3}} F_{\mu\nu} F^{\mu\nu} + M_{KK}^2 \left(1+Z^2\right) F_{\mu Z}
F^\mu_{\,\,\,Z}\right) \ , \label{5dtheory}
\ee
and the Chern-Simons term,
\be
S_{CS}= {N_c\over 24\pi^2}
\int_{R^4\times Z} {\rm Tr}\left(A F^2+ {i\over 2} A^3 F -{1\over
10}A^5\right) \ . \label{CS5d}
\ee
where $ A = A_{\mu}dx^{\mu} + A_z dZ $, and the metric is chosen to be mostly negative. From what follows, we will focus on the case with only two flavors $N_f=2$ that is sufficient to our present purposes.

The non-Abelian gauge fields $A_M(x^\mu,Z)$ contain all information about the pion, vector and
axial-vector mesons.
More precisely, the 5D gauge field $A_M$ can be mode expanded in the
$A_Z=0$ gauge as
\be
A_\mu(x,Z) = -{1\over f_\pi}\partial_\mu \pi(x) \psi_0(Z) + \sum_{n\ge 1} B_\mu^{(n)}(x)\psi_{n}(Z)+\cdots\quad,\label{mode}
\ee
where $ \psi_0(Z) = \frac{2}{\pi}\arctan(Z) $
\footnote{There is a factor 2 difference from the notation in Ref.\cite{Hong:2007kx,Hong:2007ay,Park:2008sp}.},
and $\{\psi_n(z)\}_{n\geq1}$ are ortho-normal eigenfunctions,
satisfying
\begin{align}\label{eq:EOM}
K^{1/3}\partial_Z\left(K\partial_Z\psi_n\right) = -{m^2_n\over
M_{KK}^2} \psi_n \ , \ \ \ \psi_n(Z \rightarrow \pm\infty) = 0\quad,
\end{align}
with $ K = 1 + Z^2 $ and are normalized as
\begin{align}
{\pi\over 4}{\left(f_\pi\over M_{KK}\right)^2}\int^{+\infty}_{-\infty} dZ \ K^{-1/3}\psi_n\psi_m = \delta_{mn} \ .
\end{align}
The 4D fields $\pi(x)$ and $B_\mu^{(n)}(x)$ describe pions and spin-1 mesons.
For odd integers $(2n-1)$, $B^{(2n-1)}_\mu\equiv v^{(n)}_\mu$
describe vector mesons, while $B^{(2n)}_\mu\equiv a^{(n)}_\mu$
describe axial-vector mesons \footnote{We follow the notation in
Ref.\cite{Sakai:2004cn,Sakai:2005yt} which differs from
Ref.\cite{Hong:2007kx,Hong:2007ay,Park:2008sp}}.
Inserting the field expansion from Eq.~(\ref{mode}) into the action
(\ref{5dtheory}) and (\ref{CS5d}), we get a 4D Lagrangian that
describes the dynamics of pions and tower of vector and axial-vector
mesons. The theory has a unique scale $M_{KK}$, the popular choice
of which is $M_{KK}=0.949$~GeV, chosen to reproduce the experimental
$\rho$-meson mass.


\subsection{Holographic Baryons}

Besides mesons, the present holographic model is also
able to incorporate baryons, which appear in a way quite similar to the Skyrmion in 4D \cite{Adkins:1983ya}. In the holographic 5D theory (\ref{5dtheory}), we have topological solitons along the {\it
spatial} Euclidean four dimensions $(\vec x, Z)$, whose topological charges are determined by the second Chern number of the gauge potential
\be
B={1\over 8\pi^2}\int_{R^3\times Z}\,{\rm Tr}\left(F\wedge  F \right)\quad.\label{topcharge}
\ee
In fact this topological charge is related to the usual Skyrmion number of pions upon 4D interpretation (\ref{mode}), which supports the identification of these solitons as baryons \cite{Son:2003et}.

As the energy of the baryon is minimized at $Z=0$, due to the specific form of the action (\ref{5dtheory}), it localizes
at this position. In fact, the size of this soliton baryon naively tends to shrink to zero by the same reason.
From the gauge-kinetic action (\ref{5dtheory}), it follows that the mass of the soliton ($M_B$) with $B=1$, which is localized at $Z=0$ and
has zero size would be
\be M_{B}^0 = 4\pi^2 \ \frac{\pi}{2}
\left(\frac{f_{\pi}}{M_{KK}}\right)^2 M_{KK} \ . \label{zeromass}\ee
If the size of the soliton is perturbed by $\rho$, then the increase in mass is $\delta M_B \sim \rho^2$. This suggests that it would be energetically more favorable for the soliton to stay localized at $Z=0$ with zero size.

However, the presence of the 5D Chern-Simons term (\ref{CS5d})
induces an additional electric charge to the topological charge
(\ref{topcharge}), whose electrostatic self-energy ($\sim
1/\rho^2$) compensates the shrinking tendency. As a result, the
baryon acquires a stable and finite size
\cite{Hong:2007kx,Hata:2007mb,Hong:2007ay}. Note that the soliton
profile should be much like the profile of the instanton in
Yang-Mills theory, since the equations that describe this soliton
are similar to those of the Euclidean 4D YM theory, except a
nontrivial form of the action (\ref{5dtheory}).

The soliton solutions come with a moduli space of zero modes, just like in case of the Skyrmions, and one should quantize along this moduli space. The structure of the moduli space can be easily learned from that of the instantons: we have approximate $SO(4)$
spatial rotations interlocked with the isospin $SU(2)_I$ in addition to the usual translational moduli. Size is not a moduli due
to the non-trivial $Z$ dependence of the metric, but one can also choose to quantize along the size as in Ref.~\cite{Hata:2007mb}. As a
result, one gets nucleons as lowest states of the spectrum, $\Delta$ baryons as a next excitation, and other higher excited baryon states.

One of the important lessons we learn from the well-known YM instanton solution is its purely non-Abelian nature in the field
profile with the $U(1)$ part being absent. Another lesson is that the profile has a long-ranged tail at large distance $r$ of the form
\be
A_m^a \sim -\rho^2{\bar\eta}^a_{mn}\partial_n{1\over r^2}\quad,\label{tail}
\ee
where $\rho$ is the stabilized size of the soliton baryon. This can provide important information about the coupling of baryons with
the mesons, more precisely with the 5D gauge field $A_M(x,Z=0)$ at the position of the soliton $Z=0$.
This is essentially in the same vein to the usual pion tail of the Skyrmion solution, whose strength was used by Adkins-Nappi-Witten to obtain the nucleon-pion coupling $g_{\pi NN}$ \cite{Adkins:1983ya}. Recall that the logic was to replace the {\it quantized} Skyrmions with a pointlike nucleon field $N(x)$ that has a coupling to the pions with the right strength $g_{\pi NN}$ to source the asymptotic pion tail of the original Skyrmion solution. One subtlety was that the pion tail of the classical Skyrmion solution must be replaced by the expectation value of {\it quantum} states obtained by semiclassical quantization along the zero modes. Only after that it can be mapped
to the profile sourced by the nucleon current ${\bar N}\gamma^\mu\gamma^5 N$ in the {\it nonrelativistic} limit.
Also, the quantization of Skyrmion results in a tower of higher half-integer spin and isospin states as well, with $\Delta$ baryons
of spin and isospin $3/2$ being a primary example. Through the identical procedure as for nucleons, it was also possible to
calculate nucleon-$\Delta$-pion coupling $g_{N\Delta\pi}$ \cite{Adkins:1983ya}.

The same logic may be invoked to find a coupling of {\it quantized} holographic baryons in 5D to gauge field $A_M(x^\mu,Z)$ that encodes
pions and a tower of spin-1 vector mesons via (\ref{mode}). In other words, the coupling of the baryon field to the 5D gauge field $A_M$
must be such to be able to reproduce {\it quantum state expectation value} of the long-ranged tail (\ref{tail}) of the original soliton
solution for the baryon. The necessary semiclassical quantization on the zero modes is essentially identical to that of the Skyrmion,
giving us a tower of {\it nonrelativistic} spectrum of half-integer spin and isospin. By computing expectation values of the long-range
tail (\ref{tail}) over these states, one may write down effective local couplings of baryon fields to the field $A_M$ that can
reproduce these quantum averaged tails. An important point is that these couplings will involve only the non-Abelian part of $A_M$ as the
Abelian $U(1)$ component is absent in the tail (\ref{tail}).


For the lowest spin, isospin $1/2$ states corresponding to a holographic version of the usual nucleons, Refs.~\cite{Hong:2007kx,Hong:2007ay,Hong:2007dq} introduced a 5D Dirac spinor ${\cal B}$ to write down an effective Lagrangian that
encapsulates the above features. The necessary analysis for higher excited baryons was done in Ref.~\cite{Park:2008sp}. However, in 5D
it seems hard to find a fully relativistic formulation of higher spin fermions, and the effective fields and Lagrangians in
Ref.~\cite{Park:2008sp} are only nonrelativistic. In the next subsection, we will summarize this development which serves as a
basis of our subsequent analysis. Later, we will show how one can proceed to a necessary relativistic extension of their results, that
will be crucial for our purposes.


\subsection{Nonrelativistic Treatment}

In this section we summarize some of the results of Ref.~\cite{Park:2008sp}, which will be our starting point in calculating the transition form factors. As we are interested only in $N-\Delta$ transition in this work, the relevant interaction term would involve baryon fields of spin and isospin $1/2$ as well as $3/2$.
A convenient {\it nonrelativistic} notation for spin, isospin $1/2$ states is
\be
{\cal U}_\alpha^\epsilon\quad,
\ee
where $\alpha=+,-$ is a two-component nonrelativistic spinor index and $\epsilon=1,2$ is the index for the isospin doublet. A similar
notation for spin, isospin $3/2$ states is
\be
{\cal U}_{\alpha_1 \alpha_2 \alpha_3}^{\epsilon_1 \epsilon_2 \epsilon_3}\quad,
\ee
where the three indices $\alpha_i$'s and $\epsilon_i$'s must be totally symmetric to be in the $S = I = 3/2$ representation under
the spatial rotation and the isospin. These {\it nonrelativistic} baryon fields are presumed to be localized at the position $Z=0$,
where they minimize their energy. A quantum spread along $Z\neq 0$ can be shown to be a sub-leading effect in the large $N_c$ limit
that holographic QCD is based on. With the above notations, the relevant {\it nonrelativistic} $N-\Delta$ transition coupling to
the 5D gauge field $A_M=A_M^a {\tau_a\over 2}$ ($a=1,2,3$) has been obtained to be \cite{Park:2008sp}
\be
S_{N\Delta}^{5D}=
- k_{1\over 2}\left(\frac{1}{2}F_{ij}^a \epsilon_{ijk}\left({\cal U}_\alpha^\epsilon\right)^*
\left(\sigma_2\sigma_k\right)^{\beta \beta'}\left(\tau_2\tau_a\right)^{e e'}{\cal U}_{\beta\beta' \alpha}
^{e e'\epsilon}
+ F_{Zi}^a
\left({\cal U}_\alpha^\epsilon\right)^*
\left(\sigma_2\sigma_i\right)^{\beta \beta'}\left(\tau_2\tau_a\right)^{e e'}{\cal U}_{\beta\beta' \alpha}
^{e e'\epsilon}\right)\Bigg|_{Z=0},\label{ndelta}
\ee
where
\be
k_{1\over 2} = {\sqrt{3}(N_c+2)\over 4\sqrt{5}}M_{KK}^{-1} = {\sqrt{15}\over 4}M_{KK}^{-1} \ .
\ee
We have replaced $N_c$ by $N_c+2$, as was argued in Refs.~\cite{Hong:2007kx,Hong:2007ay,Amado:1986ef} to better account
for subleading effects. We will discuss this issue in more detail in Sec.~4.
Note that the Abelian $U(1)$ part is absent in the above due to the
non-Abelian nature of the soliton-baryon profile, as explained before. By inserting the mode expansion (\ref{mode}) of $A_M$ in
Eq.~(\ref{ndelta}), one can obtain nucleon-$\Delta$ couplings to the pions and a tower of spin-1 (axial) vector mesons.

For our purpose of calculating electromagnetic (EM) form factors, we
need to consider vector mesons $B^{(2n-1)}_\mu=v^{(n)}_\mu$ only, as
EM does not couple to axial vectors. Because for vector mesons
$v^{(n)}_\mu(x^\mu)$, $\psi_{(2n-1)}(Z)$ is even under $Z\to -Z$, it
is easily seen that the second term in (\ref{ndelta}) simply
vanishes, and the first piece gives us
\be
S_{N\Delta v}= -\frac{1}{2}k_{1\over 2}\sum_{n\ge 1}\psi_{(2n-1)}(0)
\left(\partial_i v^{(n)}_j-\partial_j v^{(n)}_i\right)^a
\epsilon_{ijk}
\left({\cal U}_\alpha^\epsilon\right)^*
\left(\sigma_2\sigma_k\right)^{\beta \beta'}\left(\tau_2\tau_a\right)^{e e'}{\cal U}_{\beta\beta' \alpha}
^{e e'\epsilon}\,.\label{ndeltanr}
\ee
This {\it nonrelativistic} result will be the starting point of our analysis. Notice that due to $F_{ij}^a \epsilon_{ijk}$ structure in
(\ref{ndelta}) this interaction is of ``magnetic'' type. Since $\psi_{(2n-1)}(Z)$ is an even function for the vector mesons, there
is no analogous ``electric'' type of interaction (unless there is some sub-leading asymmetric smearing in $Z$). This observation is in
agreement with the fact that the EM transition of $N\Delta$ is predominantly of magnetic ($M1$) dipole type.

It is also instructive to consider $N\Delta \pi$ coupling. In this case, since $\psi_0(Z)$ is an odd function, only the second term
survives and as a result \cite{Park:2008sp},
\be
S_{N\Delta\pi} = \frac{8k_{1\over 2}}{\pi f_{\pi}}\left(\partial_i\pi^a\right)
\left({\cal U}_\alpha^\epsilon\right)^*
\left(\sigma_2\sigma_i\right)^{\beta \beta'}\left(\tau_2\tau_a\right)^{e e'}{\cal U}_{\beta\beta' \alpha}
^{e e'\epsilon} \ .
\ee
The generalization of this interaction to the relativistic case may be important when discussing the photoproduction processes.


\section{Vector Dominance in Holographic QCD}

In the Sakai-Sugimoto model, the 5D $U(N_f)$ gauge field $A_M$
($M=0,1,2,3,Z$) contains pseudoscalar pions and a tower of
vector/axial-vector mesons upon its 4D mode expansion. It can also
include external vector potentials that couple to $U(N_f)_L \times
U(N_f)_R$ chiral symmetry currents as its non-normalizable modes
near $Z\to \pm \infty$ boundaries. In the $A_Z=0$ gauge (leading
order in fields), the mode expansion reads as
\cite{Sakai:2004cn,Sakai:2005yt}
%
%
\be
A_\mu(x,Z)=\left(-{1\over f_\pi}\partial_\mu \pi(x)+{\cal A}_\mu(x)\right) \psi_0(Z)
+ {\cal V}_\mu(x) + \sum_{n\ge 1} B_\mu^{(n)}(x)\psi_{(n)}(Z)+\cdots\quad,\label{expand}
\ee
where ${\cal V}={1\over 2}\left(A_L+A_R\right)$ and ${\cal
A}={1\over 2}\left(A_L-A_R\right)$ are the external vector and
axial-vector potentials. By looking at how the model responds to
these external potentials, one can study various form factors of
chiral symmetry currents.
Electroweak form factors of the QCD sector would be of particular interest for applications, see e.g. Ref.~\cite{Gazit:2008gz}.

The electromagnetic vector potential $A^{EM}_\mu$ can be thought of as an external potential probing the QCD sector by
\be
{\cal V}_\mu = e \cdot \left(\begin{array}{cc} {2\over 3} &0 \\0 & -{1\over 3} \end{array}\right)A^{EM}_\mu
\quad,\quad {\cal A}_\mu =0 \quad.\label{em}
\ee
As the axial part ${\cal A}_\mu$ is absent, the electromagnetic external potential will couple only to the vector mesons
$B^{(2n-1)}_\mu \equiv v^{(n)}_\mu$, which allows one to neglect axial-vector mesons $B^{(2n)}_\mu\equiv a^{(n)}_\mu$ in the above
expansion (\ref{expand}). We will see how ${\cal V}_\mu$ interacts with the system in the following, neglecting the
axial part, which results in a feature quite similar to vector dominance with a tower of vector mesons $v^{(n)}_\mu$.

Keeping only the vector part, and using Eq.~(\ref{eq:EOM}), we get
\be
\left(1+Z^2\right)^{1\over 3}\partial_Z\left[\left(1+Z^2\right) \partial_Z \psi_{(2n-1)}\right]
=-{m^2_{v^n}\over M_{KK}^2} \psi_{(2n-1)}\quad,
\ee
as well as the orthonormality condition:
\be
{\pi\over 4}{\left(f_\pi\over M_{KK}\right)^2}
\int_{-\infty}^{+\infty} dZ\,\left(1+Z^2 \right)^{-{1\over 3}}\psi_{(2n-1)}(Z)\psi_{(2m-1)}(Z)=\delta_{nm}\quad .
\ee
Taking all of these into account, the action (\ref{5dtheory}) for $A_M$ reduces to a 4D action
\be
S_{4D}=\int d^4 x \, \left[ {\rm Tr}\left(-{\cal F}^{\cal V}_{\mu\nu} \sum_{n\ge 1} a_{{\cal V}v^n}
F^{(n)\mu\nu}\right)
+\sum_{n\ge 1}{\rm Tr}\left(-{1\over 2}F_{\mu\nu}^{(n)}F^{(n)\mu\nu} +m^2_{v^n} v_\mu^{(n)} v^{(n)\mu}
\right)\right]\quad,
\ee
where
${\cal F}^{\cal V}_{\mu\nu}=\partial_\mu {\cal V}_\nu -\partial_\nu {\cal V}_\mu$,
$F_{\mu\nu}^{(n)}=\partial_\mu v_\nu^{(n)}-\partial_\nu v_\mu^{(n)}$
and \footnote{In Ref.\cite{Hong:2007kx,Hong:2007ay,Park:2008sp},
$a_{{\cal V}v^n}$ is written instead as $\zeta_n$, and $m_{v^n}^2$ as $m_{(2n-1)}^2$.}
\be
a_{{\cal V} v^n}=
{\pi\over 4}{\left(f_\pi\over M_{KK}\right)^2}
\int_{-\infty}^{+\infty} dZ\,\left(1+Z^2 \right)^{-{1\over 3}}\psi_{(2n-1)}(Z)\quad.\label{avn}
\ee
The second term in this action describes massive vector mesons, while the first piece represents
a kinetic mixing between the external vector potential ${\cal V}_\mu$ and the massive vector mesons $v^{(n)}_\mu$.
It is more convenient to diagonalize the kinetic terms by shifting the vector meson fields as
\be
v^{(n)}_\mu = {\tilde v}_\mu^{(n)} - a_{{\cal V}v^n} {\cal V}_\mu \quad,
\ee
which transforms the action to
\be
S_{4D}=\int d^4 x \, \sum_{n\ge 1}{\rm Tr}\left(-{1\over 2}{\tilde F}_{\mu\nu}^{(n)}
{\tilde F}^{(n)\mu\nu} +m^2_{v^n}{\tilde v}_\mu^{(n)} {\tilde v}^{(n)\mu}-2 m^2_{v^n}a_{{\cal V}v^n}
{\tilde v}^{(n)}_\mu
{\cal V}^\mu
\right)\quad,\label{mix}
\ee
in terms of ${\tilde v}^{(n)}_\mu$ fields, up to an additive
renormalization of ${\cal V}_\mu$ kinetic terms which are divergent
anyway. In the ${\tilde v}^{(n)}_\mu$ basis, the mixing to ${\cal
V}_\mu$ is independent on the momentum transfer, which will make the
summation over $n$ of Feynman diagrams more convergent as we will
see later. This is the usefulness of this new basis, although any
final results must be independent of whether we work in
$v^{(n)}_\mu$ or ${\tilde v}^{(n)}_\mu$ basis.

Another advantage in using this new basis ${\tilde v}^{(n)}_\mu$ is in a manifest presence of the holographic vector meson dominance feature.
Note that the expansion of $A_M$ in (\ref{expand}) including only the vector part becomes in the new basis
\be
A_\mu(x,Z)& =& {\cal V}_\mu(x) +\sum_{n\ge 1}v^{(n)}_\mu(x)\psi_{(2n-1)}(Z)\nonumber \\
&=&\left(1-\sum_{n\ge 1}a_{{\cal V}v^n} \psi_{(2n-1)}(Z)\right) {\cal V}_\mu(x)
+\sum_{n\ge 1}{\tilde v}^{(n)}_\mu(x)\psi_{(2n-1)}(Z)\quad.
\ee
Using the completeness relation of the eigenfunctions for the vectorlike quantity
\be {\pi\over 4}{\left(f_\pi\over M_{KK}\right)^2} \sum_{n\ge
1}\left(1+Z^2 \right)^{-{1\over 3}}\psi_{(2n-1)}(Z)\psi_{(2n-1)}(Z')
=\delta(Z-Z')\quad,\label{completeness} \ee
and the definition (\ref{avn}) of $a_{{\cal V}v^n} $, we can easily derive a {\it sum rule}:
\be
\sum_{n\ge 1}a_{{\cal V}v^n} \psi_{(2n-1)}(Z)=1\quad,\label{sumrule}
\ee
which drastically simplifies the expansion
\be
A_\mu(x,Z)=\sum_{n\ge 1}{\tilde v}^{(n)}_\mu(x)\psi_{(2n-1)}(Z),\label{VMD}
\ee
without any remnant of the external vector potential ${\cal V}_\mu$. In other words, in the ${\tilde v}^{(n)}_\mu$ basis the
external vector potential ${\cal V}_\mu$ does not interact directly with the system through $A_M$, but only through
momentum-independent mixings with the vector mesons ${\tilde v}^{(n)}_\mu$ via (\ref{mix}). Any interaction of the system with
the external vector potential is completely mediated by tree-level exchanges of the massive vector mesons ${\tilde v}^{(n)}_\mu$.
Moreover, one can define vector meson decay constants as follows:
\be
\langle 0|J_{V\mu}^a(0)|v^{(n),b}\rangle = g_{v^n}\delta^{ab}\epsilon_{\mu} \ ,  \ \ \ g_{v^n} \equiv m^2_{v^n}a_{{\cal V}v^n} \ .
\ee
Therefore, the previous interaction term (\ref{ndeltanr}) of $N-\Delta$ with vector mesons can be easily generalized to the case
of having an external vector potential ${\cal V}_\mu$, by simply replacing $v^{(n)}_\mu$ with ${\tilde v}^{(n)}_\mu$:
\be
S_{N\Delta{\tilde  v}}= - \frac{1}{2}k_{1\over 2}\sum_{n\ge 1}\psi_{(2n-1)}(0)
\left(\partial_i {\tilde v}^{(n)}_j-\partial_j {\tilde v}^{(n)}_i\right)^a
\epsilon_{ijk}
\left({\cal U}_\alpha^\epsilon\right)^*
\left(\sigma_2\sigma_k\right)^{\beta \beta'}\left(\tau_2\tau_a\right)^{e e'}{\cal U}_{\beta\beta' \alpha}
^{e e'\epsilon} \ ,\label{NRver}
\ee
without any further change. The electromagnetic $N-\Delta$ transition form factors will be described by tree-level Feynman diagrams, where the external field, ${\cal V}_\mu$, couples to transition vertex only through the exchange of vector mesons ${\tilde
v}^{(n)}_\mu$.

\section{Other Approaches}

We should point out that there are other similar approaches in the literature, exploring chiral symmetry currents
from the soliton solution itself \cite{Nawa:2006gv,Hata:2007mb,Hata:2008xc,Hashimoto:2008zw,Kim:2008pw,Seki:2008mu}.
The only difference can be traced back to how we treat the {\it nonrelativistic} baryon wavefunction along the $Z$ direction,
which was taken to be the $\delta$-function in the above, to give the $\psi_{(2n-1)}(0)$ factor.
Although this is a right thing to do in a strict large $N_c$ limit sense, Refs.~\cite{Hata:2007mb,Hata:2008xc,Hashimoto:2008zw,Kim:2008pw,Seki:2008mu} went one step further to better approximate
the baryon wavefunctions, which corresponds to including a subleading effect in the large $N_c$ limit.
This modification of the baryon wavefunctions will have its effects only on the factor $\psi_{(2n-1)}(0)$.

In Ref.~\cite{Hata:2007mb,Hashimoto:2008zw}, the motion along $Z$ was approximated by harmonic oscillator. For our purposes, it is sufficient to take the ground state wavefunction:
\be
\Psi_B^{n_Z=0}(Z)=\left(2a\over\pi\right)^{1\over 4}\,\, e^{-a Z^2}\quad,
\ee
where $a={2\pi^3\over\sqrt{6}}\left(f_\pi\over M_{KK}\right)^2 \approx 0.240$.
Accordingly, we replace $\psi_{(2n-1)}(0)$ by
\be
\langle\psi_{(2n-1)}(Z)\rangle = {\int_{-\infty}^{+\infty} dZ \,\, \psi_{(2n-1)}(Z)\,e^{-2a Z^2}
\over \int_{-\infty}^{+\infty} dZ \,\, e^{-2a Z^2}}\quad.
\ee
In fact Refs.~\cite{Hata:2007mb,Hashimoto:2008zw} also treated the size of soliton-baryon, $\rho$, as a quantum mechanical modulus,
and any quantity that involves $\rho$ should also be averaged over the resulting wavefunction on $\rho$.
It can be shown that the coefficient $k_{1\over 2}$ in (\ref{ndelta}) is proportional to $\rho^2$, as the long-ranged tail of
the soliton-baryon that this term is based on is linear in $\rho^2$, which can be seen in (\ref{tail}).
The resulting quantum average increases $\rho^2$ and hence $k_{1\over 2}$ by a factor of
\be
{\sqrt{5}+2\sqrt{5+N_c^2}\over 2 N_c} \approx 1.62 \quad.
\ee
Recall that the earlier expression for
$k_{1\over 2}={\sqrt{3}(N_c+2) \over 4\sqrt{5}}M_{KK}^{-1} = {\sqrt{15}\over 4}M_{KK}^{-1}$
involves a shift $N_c \to N_c + 2$ that effectively accounts a subleading effect in the large $N_c$ limit \cite{Amado:1986ef}.
This observation was used in the analysis of Refs.~\cite{Hong:2007kx,Hong:2007ay,Hong:2007dq}.
This shift corresponds to the increase of $k_{1\over 2}$ by a factor of ${5\over 3}\approx 1.66$ relative to its classical value,
which happens to be very close to the above increase by the quantum wavefunction on $\rho$.
This in fact explains the {\it fortunate} numerical agreements between Refs.~\cite{Hong:2007kx,Hong:2007ay,Hong:2007dq}
and Ref.~\cite{Hashimoto:2008zw}. Although it doesn't make much difference to replace our previous value
$k_{1\over 2}={\sqrt{15}\over 4}M_{KK}^{-1}\approx 0.968 M_{KK}^{-1}$
with the value from the analysis in Ref.~\cite{Hashimoto:2008zw},
\be
k_{1\over 2}=1.62\times {\sqrt{3} N_c \over 4\sqrt{5}}M_{KK}^{-1}\approx 0.941 M_{KK}^{-1}\quad,
\ee
we will choose the latter for consistency when we discuss the results by the approach of Ref.~\cite{Hashimoto:2008zw}.

As the soliton-baryon size $\rho$ is given by
\be
\rho^2 ={N_c\over \pi^3}\sqrt{3\over 10}\left(M_{KK}\over f_\pi\right)^2\approx (2.364)^2 \quad,
\ee
which is {\it numerically bigger} than the average size of the quantum wavefunction in Ref.~\cite{Hashimoto:2008zw},
\be
Z_{av}=\sqrt{\langle Z^2 \rangle}={1\over 2}\sqrt{1\over a} \approx 1.021\quad,
\ee
the quantum spread over $Z$ seems numerically subdominant to the initial size effect of the baryons.
This seems to be a problem in this approach.

For completeness, we will present both results we obtain using the approaches in Refs.~\cite{Hata:2007mb,Hashimoto:2008zw} and the previous one with the factor $\psi_{(2n-1)}(0)$
based on Ref.~\cite{Park:2008sp}. We refer the former as the Type II and the latter as the Type I model.
In summary,
\be
{\rm Type\,I\,}&:&\quad\quad \psi_{(2n-1)}(0)\nonumber\\
{\rm Type\,II\,}&:&\quad\quad \psi_{(2n-1)}(0)\to \langle \psi_{(2n-1)}(Z) \rangle \nonumber
\ee
As a final comment, the two types of approaches we consider share one common feature that seems to be
universal.
Because of the identity (\ref{sumrule})
\be
\sum_{n\ge 1}a_{{\cal V}v^n} \psi_{(2n-1)}(Z)=1\quad,
\ee
for {\it any} $Z$, one can easily see that the {\it sum rule}
\be
\sum_{n\ge 1}a_{{\cal V}v^n} \langle \psi_{(2n-1)}(Z)\rangle =1\quad,
\ee
holds whatever approximation we use for $\langle \psi_{(2n-1)}(Z)\rangle$. This will give us a universal result for the zero-momentum
limit of the form factor, that is a one robust prediction of the model without referring to a specific type of approach.


\section{Relativistic Generalization}

As was discussed earlier, in the holographic model, the external electromagnetic field is carried by the vector mesons. The linear coupling of the electromagnetic potential with these vector mesons is given by the last term in the interaction
Lagrangian (\ref{mix}), where the external vector field is expressed through the electromagnetic potential as in Eq.~(\ref{em}).
Therefore, the interaction of the electromagnetic field with the hadrons can only occur through the intermediate vector meson exchange.
This feature of ``holographic vector meson dominance,'' given by (\ref{VMD}), is a relativistic concept, since it emerges from
the manifestly relativistic 5D gauge theory. On the other hand, the nucleon-$\Delta$-${\tilde v}^{(n)}$
interaction vertex, given by Eq.~(\ref{NRver}), is {\it nonrelativistic}.
To have a more consistent framework, we have to find a relativistic generalization of this vertex (\ref{NRver}). Although, we are interested in low momentum transfers, where the nonperturbative effects are dominant, there is no clear separation of relativistic and nonrelativistic effects.
Moreover, for momentum transfers larger than about $2 \ \GeV^2 $, the relativistic effects may not be negligible.
With this generalization in hand, we can find the relativistic nucleon-$\Delta$-${\tilde v}^{(n)}$ transition form factors by simply summing over all tree-level Feynman diagrams involving ${\tilde v}^{(n)}_\mu$ meson exchanges.

A difficulty in Ref.~\cite{Park:2008sp} for a relativistic formulation was the absence of a relativistic 5D formalism of high spin fermions, notably for spin $3/2$ fermions corresponding to a holographic description of $\Delta$ baryons. This difficulty might
have a chance to be overcome in the future, but there is one point we can make at present without any regard to details
of a solution: the resulting 4D theory, after integrating over $Z$, must be reduced to a 4D relativistic theory for a spin $3/2$
field that describes the $\Delta$ resonances.
Moreover,  for consistency, the {\it nonrelativistic} limit of this 4D theory should precisely reproduce the previous {\it
nonrelativistic} result (\ref{NRver}).

In the {\it nonrelativistic} treatment of Ref.~\cite{Park:2008sp}, a $\delta$-function localization of the baryon
wavefunctions to $Z=0$ was assumed, as a leading approximation to the large $N_c$ limit, where the baryons become infinitely heavy.
This seems to indicate an intricate intertwining of a relativistic generalization and an inclusion of sub-leading effects of the large $N_c$ limit: if we take the large $N_c$ limit first, then the baryons should be treated nonrelativistically.

We propose to take a different path instead. We first impose a relativistic formulation before looking at the leading large $N_c$
effects. After integrating over $Z$, a presumed 4D relativistic theory will be a theory of spin $3/2$ $\Delta$ baryons, for which we have a consistent relativistic description in terms of the Rarita-Schwinger formalism. Whatever formalism for a 4D relativistic spin $3/2$ field we would have from the perspective of a presently unknown 5D relativistic formulation, it would be equivalent to the Rarita-Schwinger field at
the end.
As we take the large $N_c$ limit for the parameters in the resulting
relativistic theory, we expect that in the nonrelativistic limit
the results must agree with those of Ref.~\cite{Park:2008sp}. This
expectation relies on the assumption of validity of exchanging the
order of the two limits. This way one will be able to fix the
parameters in the 4D relativistic theory of Rarita-Schwinger
$\Delta$ baryons {\it in the large $N_c$ limit}.

A relativistic generalization of the nucleon-$\Delta$-${\tilde v}^{(n)}$ vertex (\ref{NRver}) of interest in terms of Dirac
spinor nucleons and Rarita-Schwinger $\Delta$ baryons can be determined within our proposal, which will be the subject of the
next sections.

\subsection{Basics of Rarita-Schwinger Spin $3\over 2$ Field}

The $\Delta(1232)$ is a spin-$3/2$ resonance. Therefore its spin content can be described in terms of a Rarita-Schwinger (RS) field
\cite{Rarita:1941mf}: $\Psi_\mu^{(\sigma)}$, where $\mu$ is the vector and $\sigma$ the spinor index (the latter index is omitted in
the following).
Here, we briefly summarize the relevant relativistic formalism for this 4D RS field.

The free and massive RS field obeys the Dirac equation, supplemented with the auxiliary conditions (constraints):
\begin{align}\label{EOMconstr}
(i\sla{\partial} - M)\Psi_\mu  = 0 \ , \ \ \ \partial^{\mu}\Psi_{\mu} = 0 \ , \ \ \ \gamma^{\mu} \Psi_{\mu} = 0 \ .
\end{align}
The constraints ensure that the number of independent components of the vector-spinor field is reduced to the physical number of spin
degrees of freedom.
In the interacting theory, the coupling of the RS field must be compatible with the free theory construction in order to preserve
the physical spin degrees of freedom, otherwise one ends up with unphysical degrees of freedom with negative-norm
states~\cite{Johnson:1960vt,Hagen:1972ea} or superluminal (acausal) modes~\cite{Velo:1969bt,Singh:1973gq}.
The proposals for consistent spin-$3/2$ couplings were given e.g. in Ref.~\cite{Hagen:1982ez}.
Although it can be a subtle issue to discuss consistency of a quantum theory of interacting RS field, we will not be concerned
about this here, since the Feynman diagrams, that are required to calculate form factors we are interested in, are tree-level diagrams
and involve intermediate vector meson exchanges.
In fact, we are only interested in the kinematic description of relativistic spin-$3\over 2$ particle and its pointlike tree-level
interactions with massive vector mesons ${\tilde v}^{(n)}_\mu$, such that these interaction vertices reduce to our previous expression
(\ref{NRver}) in the nonrelativistic limit.

We will work in the conventions, where the $\gamma$ matrices are
\be
\gamma^0=\left(\begin{array}{cc}1 & 0\\0&
-1\end{array}\right)\quad,\quad \gamma^i=\left(\begin{array}{cc}0 &
\sigma_i\\ -\sigma_i & 0\end{array}\right)\quad,\quad
\gamma^5=\left(\begin{array}{cc}0 & 1\\1& 0
\end{array}\right)\quad.
\ee
Since we are interested in the nonrelativistic limit for the $\Delta$ resonances,
consider a specific momentum state of $\partial_\mu = -i p_\mu$, which has the following restframe
components $p_0=E=M$ and $p_i=0$ ($i=1,2,3$).
Then solving equations of motion with constraints, one finds $\Psi_0=0$ and
\be
\Psi_i=\left(\begin{array}{c} {\cal U}_i \\
0\end{array}\right)\quad,
\ee
with three two-component spinors ${\cal U}_i$ ($i=1,2,3$) satisfying an important constraint,
\be
\sum_{i=1}^3 \sigma_i\, {\cal U}_i =0 \quad.\label{rschcon}
\ee
In fact, ${\cal U}_i$ ($i=1,2,3$) with the constraint (\ref{rschcon}) is an unconventional way of describing usual nonrelativistic spin
$s={3\over 2}$ states of $SO(3)$ rotation group. First note that the independent number of components is indeed $2\times 3-2=4$ as in the
case of spin $s=3/2$ representation. It is not difficult to find the similarity transformation between the ${\cal U}_i$
representation and the usual representation with $|s={3\over 2},s_z\rangle$ basis.
Writing
\be
{\cal U}_i=\left(\begin{array}{c}
a_i\\b_i\end{array}\right)\quad,
\ee
with complex numbers $a_i$ and $b_i$, one finds
\ba
a_1 &=& \sqrt{1\over 2}\,\, \Big|{3\over
2},{3\over 2}\Big\rangle + \sqrt{1\over 6}\,\, \Big|{3\over
2},-{1\over 2}\Big\rangle\quad,\quad b_1 = \sqrt{1\over 2}\,\,
\Big|{3\over 2},-{3\over 2}\Big\rangle + \sqrt{1\over 6}\,\,
\Big|{3\over 2},{1\over 2}\Big\rangle\quad,\nonumber\\
a_2 &=& i\sqrt{1\over 2}\,\, \Big|{3\over 2},{3\over 2}\Big\rangle -
i\sqrt{1\over 6}\,\, \Big|{3\over 2},-{1\over 2}\Big\rangle
\quad,\quad b_2 = -i\sqrt{1\over 2}\, \,\Big|{3\over 2},-{3\over
2}\Big\rangle + i\sqrt{1\over 6}\,\,
\Big|{3\over 2},{1\over 2}\Big\rangle\quad,\nonumber\\
a_3&=& -\sqrt{2\over 3}\,\,\Big|{3\over 2},{1\over
2}\Big\rangle\quad,\quad b_3= \sqrt{2\over 3}\,\,\Big|{3\over
2},-{1\over 2}\Big\rangle\quad.\label{simtrans1}
\ea
Recall that in our expression (\ref{NRver}), we used yet another form of spin $s={3\over 2}$ representation: a
three-indexed objects ${\cal U}_{\alpha_1\alpha_2\alpha_3}$ totally symmetric under permutations of $\alpha_i=+,-$. It is easy to relate
this representation with the standard $|s={3\over 2},s_z\rangle$ representation,
\ba
{\cal U}_{+++}&=&\Big|{3\over 2},{3\over 2}\Big\rangle\quad,\nonumber\\
{\cal U}_{++-}&=&{\cal U}_{+-+}={\cal U}_{-++}=\sqrt{1\over 3}\,\,
\Big|{3\over 2},{1\over 2}\Big\rangle\quad,\nonumber\\
{\cal U}_{--+}&=&{\cal U}_{-+-}={\cal U}_{+--}=-\sqrt{1\over 3}\,\,
\Big|{3\over 2},-{1\over 2}\Big\rangle\quad,\nonumber\\
{\cal U}_{---}&=&-\Big|{3\over 2},-{3\over
2}\Big\rangle\quad,\label{simtrans2}
\ea
where sign conventions are chosen simply for later convenience. With the above (\ref{simtrans1}) and (\ref{simtrans2}), one can now easily
translate the nonrelativistic limit of the RS field and the three-indexed object ${\cal U}_{\alpha_1\alpha_2\alpha_3}$ that
was used in (\ref{NRver}). One identity that we will need specifically is
\be \left(\sigma_{[i} \,{\cal U}_{j]}\right)_\alpha=
{1\over 2\sqrt{2}}\epsilon_{ijk}
\left(\sigma_2\sigma_k\right)^{\beta\beta'}{\cal
U}_{\beta\beta'\alpha}\quad,\label{impid}
\ee
with $[i,j]={1\over 2}(ij-ji)$, that is straightforward to check using (\ref{simtrans1}) and (\ref{simtrans2}).


\subsection{Relativistic $N\Delta{\tilde v}^{(n)}$ Vertex}

With the gadget in the previous section, we can now find the
relativistic $N\Delta{\tilde v}^{(n)}$ vertex, that generalizes the
nonrelativistic expression (\ref{NRver}). Note that while we have
to generalize the space rotation indices (lower indices) into a
relativistic Rarita-Schwinger field, we should still keep the
isospin indices (upper indices) as they are in (\ref{NRver}). To
relate to the more conventional notation of $\Delta$-baryons, one
can simply substitute
\ba
{\cal U}^{111}&=&\Delta^{++}\quad,\quad {\cal U}^{222}=\Delta^{-}\quad,\nonumber\\
{\cal U}^{112}&=&{\cal U}^{121}={\cal U}^{211}=\sqrt{1\over 3}\,\,\Delta^{+}\quad,\nonumber\\
{\cal U}^{221}&=&{\cal U}^{212}={\cal U}^{122}=\sqrt{1\over 3}\,\,\Delta^{0}\quad.\label{condelta}
\ea
Since we don't need to modify the isospin structure in (\ref{NRver}), we will temporarily omit it,
focusing only on the relativistic generalization of the spacetime part.

Up to equations of motion, there are four possible forms of Lorentz invariant coupling
between nucleon Dirac spinor $N$, Rarita-Schwinger $\Delta$-baryon field $\Psi_\mu$, and the $n$'th massive vector mesons
$F^{(n)}_{\mu\nu}=\partial_\mu {\tilde v}^{(n)}_\nu-\partial_\nu {\tilde v}^{(n)}_\mu$:
\ba\label{terms}
(1)&& {\bar N}\,F^{(n)}_{\mu\nu}\,\gamma^\mu \,\Psi^\nu \ + \ {\rm H.c.} \nonumber\\
(2) && {\bar N}\,F^{(n)}_{\mu\nu}\,\gamma^\mu \gamma^5\, \Psi^\nu \ + \ {\rm H.c.} \nonumber\\
(3) && {\bar N}\,F^{(n)}_{\mu\nu}\,\gamma^{\mu\nu\rho}\, \Psi_\rho \ + \ {\rm H.c.} \nonumber\\
(4) && {\bar N}\,F^{(n)}_{\mu\nu}\,\gamma^{\mu\nu\rho}\gamma^5 \,\Psi_\rho \ + \ {\rm H.c.}
\ea
However, using the constraint $\gamma^\mu \Psi_\mu=0$, one can easily show that (3) and (4) are
equivalent to (1) and (2).

If we take the nonrelativistic limit as is done in the previous section, the spinors reduce to nonrelativistic two-component
spinors as
\be
N=\left(\begin{array}{c}{\cal U} \\
0\end{array}\right)\quad,\quad \Psi_i=\left(\begin{array}{c}{\cal
U}_{i}\\0\end{array}\right)\quad,\quad \Psi_0=0\quad,
\ee
and using the explicit form of the $\gamma$ matrices, one easily checks that (1) does not lead to any
nonrelativistic coupling as it couples particles to antiparticles, while (2) becomes
\be
-\left({\cal
U}_\alpha\right)^*\left(\partial_i {\tilde v}_j^{(n)}-
\partial_j {\tilde v}_i^{(n)}\right)\left(\sigma_{[i}\,{\cal U}_{j]}\right)_\alpha\quad.
\ee
From the important identity (\ref{impid}) that we derived in the previous section, this reduces to
\be -{1\over 2\sqrt{2}}
\left({\cal U}_\alpha\right)^*\left(\partial_i {\tilde v}_j^{(n)}-
\partial_j {\tilde v}_i^{(n)}\right)\epsilon_{ijk}\left(\sigma_2\sigma_k\right)^{\beta\beta'}
{\cal U}_{\beta\beta'\alpha}\quad,
\ee
which recovers precisely the spacetime index structure of our nonrelativistic coupling (\ref{NRver}).

The upshot is that the following relativistic operator,
\be
S_{N\Delta{\tilde  v}}^{\rm rel}={\sqrt{30}\over 4}
M_{KK}^{-1}\sum_{n\ge 1}\psi_{(2n-1)}(0) F_{\mu\nu}^{(n)a}
\left(\tau_2\tau_a\right)^{ee'} {\bar
N}^\epsilon\,\gamma^\mu\gamma^5\,\left(\Psi^\nu\right)^{e
e'\epsilon} + \ {\rm H.c.} \quad,
\ee
is the correct relativistic form of our nonrelativistic nucleon-$\Delta$-${\tilde v}^{(n)}$ vertex
(\ref{NRver}), where
%
$F_{\mu\nu}^{(n)a}\equiv \partial_\mu
{\tilde v}^{(n)a}_\nu-\partial_\nu {\tilde
v}^{(n)a}_\mu$,
%
and the upper indices $a$, $e$, $e'$ and $\epsilon$ represent isospin indices. With the help of
(\ref{condelta}), one can also write the final result in terms of the conventional notation of $\Delta$-baryons
($\Delta^{++},\Delta^+,\Delta^0,\Delta^-$) and the nucleons ($p,n$):
\ba
S_{N\Delta{\tilde  v}}^{\rm rel}&=& i{\sqrt{30}\over 4}
M_{KK}^{-1}\sum_{n\ge 1}\psi_{(2n-1)}(0)
F^{(n)+}_{\mu\nu}\left(\sqrt{2\over 3}\,{\bar p}\,
\gamma^\mu\gamma^5\left( \Delta^0\right)^\nu+\sqrt{2}\, {\bar n}\,
\gamma^\mu\gamma^5\left(\Delta^-\right)^\nu\right)\nonumber\\
&-& i{\sqrt{30}\over 4} M_{KK}^{-1}\sum_{n\ge 1}\psi_{(2n-1)}(0)
F^{(n)-}_{\mu\nu}\left(\sqrt{2}\,{\bar p}\, \gamma^\mu\gamma^5\left(
\Delta^{++}\right)^\nu+\sqrt{2\over 3}\, {\bar n}\,
\gamma^\mu\gamma^5\left(\Delta^+\right)^\nu\right)\nonumber\\
&+& i{\sqrt{30}\over 4} M_{KK}^{-1}\sum_{n\ge 1}\psi_{(2n-1)}(0)
F^{(n)0}_{\mu\nu}\Bigg(\sqrt{4\over 3}\,{\bar p}\,
\gamma^\mu\gamma^5\left( \Delta^+\right)^\nu+\sqrt{4\over 3}\,{\bar
n}\,
\gamma^\mu\gamma^5\left(\Delta^0\right)^\nu\Bigg) + \ {\rm H.c.} \ ,\label{finalresult}
\ea
where
\be
F^{(n)\pm}_{\mu\nu}={1\over\sqrt{2}}\left(F_{\mu\nu}^{(n)1}\mp i
F_{\mu\nu}^{(n)2}\right)\quad,\quad
F_{\mu\nu}^{(n)0}=F_{\mu\nu}^{(n)3}
\ee
are the $n$th vector meson fields in the EM charge basis, and $\Delta_\mu$ are the Rarita-Schwinger fields for the $\Delta$-baryons. Equation (\ref{finalresult}) is the final form of the sought-for relativistic couplings between nucleons, $\Delta$-baryons, and the massive vector mesons ${\tilde v}^{(n)}_\mu$.


\section{Transition Form Factors}


\subsection{Definitions}

The  $\gamma^* N \to \Delta$ transition is described by the matrix
element of the electromagnetic current $J^{\rm EM}_{\mu}$ between the nucleon state with momentum $p$ and the $\Delta$ with
momentum $ p'$. It can be written as
\begin{align}\label{Ndelta}
\langle \Delta(p') \mid J^{\rm EM}_{\mu}(0) \mid N(p)\rangle =
e \ \overline{\Psi}_{\beta}(p')\Gamma_{\beta\mu}\gamma_5 N(p) \ ,
\end{align}
where $N(p)$ and $\Psi_\beta(p')$ describe nucleon and delta, respectively.
The conservation of the electromagnetic current implies $q_{\beta}\Gamma_{\beta\mu} = 0$, where $ q = p'-p$ is the photon momentum transfer. For virtual photons ($q^2 \neq 0$),
the decomposition of the vertex function can be expressed in terms of three
independent scalar form factors $G_i(Q^2)$ with $Q^2 =-q^2$:
\begin{align}
\Gamma_{\beta\mu} = G_{1}(Q^{2})\big[q_{\beta}\gamma_{\mu} - {\sla q} g_{\beta\mu}\big]
%
%
+ G_{2}(Q^{2})\big[q_{\beta}P_{\mu} - \left(q P\right) g_{\beta\mu}\big]
%
%
+ G_{3}(Q^{2})\big[q_{\beta}q_{\mu} - q^{2}g_{\beta\mu}\big] \ ,
\end{align}
where $P = (p+p')/2$. Following \cite{Jones:1972ky},  one can also define the magnetic
dipole $G_{M}$, electric quadrupole $G_{E}$, and Coulomb
quadrupole $G_{C}$ form factors in terms of $G_{1}$, $G_{2}$, $G_{3}$ as follows:
\begin{align}\label{MEQ}
G_{M}(Q^{2}) &=
\frac{m_N}{3(m_N + m_{\Delta})}\Bigg[((3m_{\Delta} +
m_N)(m_{\Delta} + m_N) + Q^{2})\frac{G_{1}(Q^{2})}{m_{\Delta}}
%
%
+(m_{\Delta}^{2}-m_N^{2})G_{2}(Q^{2})-2Q^{2}G_{3}(Q^{2})\Bigg],
\nonumber \\[7pt]
G_{E}(Q^{2}) &= \frac{m_N}{3(m_N + m_{\Delta})}\Bigg[(m_{\Delta}^{2} - m_N^{2} - Q^{2})\frac{G_{1}(Q^{2})}{m_{\Delta}}
%
%
+ (m_{\Delta}^{2} - m_N^{2})G_{2}(Q^{2}) - 2Q^{2}G_{3}(Q^{2})\Bigg] \ ,
\nonumber \\[7pt]
G_{C}(Q^{2}) &= \frac{2m_N}{3(m_{\Delta} + m_N)}\Bigg[2m_{\Delta}G_{1}(Q^{2})
+ \frac{1}{2}(3m_{\Delta}^{2} + m_N^{2} + Q^{2})G_{2}(Q^{2})
%
%
%
+ (m_{\Delta}^{2} - m_N^{2} - Q^{2})G_{3}(Q^{2})\Bigg] \ .
\end{align}
We can also define the ratios $R_{EM}$ and $R_{SM}$
(see e.g. \cite{Buchmann:2004ia,Jones:1972ky,Caia:2004pm,Pascalutsa:2005ts}) that are often used in the experimental papers:
\begin{align}\label{Ratios}
R_{EM}(Q^2) & = \frac{E2(Q^{2})}{M1(Q^{2})} = - \frac{G_E(Q^2)}{G_M(Q^2)} \quad, \\[7pt]
R_{SM}(Q^2) & = \frac{C2(Q^{2})}{M1(Q^{2})} = -\sqrt{Q^2 +\frac{(m_{\Delta}^2- m_P^2-Q^2)^2}{4 m_{\Delta}^2}}\frac{1}{2 m_{\Delta}} \frac{G_C(Q^2)}{G_M(Q^2)} \ . \nonumber
\end{align}


\subsection{Predictions from Holographic QCD}

Adding Feynman diagrams that correspond to intermediate vector meson exchanges between the external EM current
and the $N\Delta v^{(n)}$ vertex given in Eq.~(\ref{finalresult}) (corresponding to $p \to \Delta^+ $ transition, in particular),
we will obtain the following result for the form factors:
\begin{align}
G_1(Q^2) = \sum_{n \geq 1} \frac{g_{v^n} g_{N\Delta v^n}}{Q^2 +
m^2_{v^n}} \ , \ \ \ G_2(Q^2) = G_3(Q^2) = 0 \ ,\label{summation}
\end{align}
where $g_{v^n} \equiv m_{v^n}^2 a_{\cV v^n}$ and
\begin{align}
g_{N\Delta v^n} = \sqrt{2} \ \frac{\sqrt{30}}{4M_{KK}}\langle
\psi_{(2n-1)}(Z)\rangle \ .
\end{align}
From Table I below, one can observe that for Type I model, the summation in (\ref{summation}) does not converge fast enough, while the Type
II case is sufficiently convergent.

\vspace{0.1in}

\begin{center}
\begin{tabular}[t]{|c|r|r|r|r|r|r|r|r|}
\multicolumn{8}{c}{TABLE I: Various couplings and masses for $M_{KK}=0.949$ GeV.}\\   \hline
  $ \ \ \ \ \ \ \ n$              & \ \ \ 1 \ \ \ & \ \ \ 2 \ \ \ & \ \ \ 3 \ \ \  & \ \ \ 4  \ \ \ & \ \ \ 5 \ \ \
  & \ \ \ 6 \ \ \ & \ \ \ 7 \ \ \ & \ \ \ 8 \ \ \   \\ \hline
  $ \ \ \  m_{v^n}^2 ({\rm GeV^2})$& \ 0.602 \ & \ 2.59 \ & \ 5.94  \ &  \ 10.6 \ & \ 16.7 \
  & \ 24.0 \ & \ 32.7 \ & \ 42.8 \    \\
 $ \ \ \  \ \ \ g_{v^n}({\rm GeV^2})$         &  \ 0.164 \         &  \  -0.707 \       & \  1.615 \
     & \  -2.884 \  & \ 4.508 \  & \  -6.484\
 & \ 8.869 \  & \ -11.58  \      \\
  $ \ \ \ \ \ \ g_{N\Delta v^n}({\rm GeV^{-1}})$ (I)  & \ 14.12 \          & \ 12.88 \           & \ 12.60 \  & \
  12.51 \ & \ 12.46 \  & \ 12.44  \    & \ 12.43 \  & \ 12.42 \  \\
 $ \ \ \ \ \ \ g_{N\Delta v^n}({\rm GeV^{-1}})$ (II)  & \ 11.84 \           & \ 5.512 \          & \ 0.585   \
          & \ -1.481 \              & \ -1.101 \  & \ -0.038 \ & \ 0.407 \ &  \ 0.196 \      \\
 \hline\end{tabular}
\end{center}

\vspace{0.1in}

The reason is that the completeness relation (\ref{completeness}) that we used before to
derive vector dominance is valid only with integration and not quite true pointwise, similar to the Gibbs' phenomenon in Fourier
transform theory. In these cases, the following expression
\be
 \sum_{n \geq 1} \frac{g_{v^n} g_{N\Delta v^n}}{Q^2 +
m^2_{v^n}}= \sum_{n \geq 1} \frac{g_{v^n} g_{N\Delta v^n}}{
m^2_{v^n}} -\sum_{n \geq 1} \frac{g_{v^n} g_{N\Delta v^n}
Q^2}{m^2_{v^n}(Q^2 + m^2_{v^n})} =  \sqrt{2} \
\frac{\sqrt{30}}{4M_{KK}}-\sum_{n \geq 1} \frac{g_{v^n} g_{N\Delta
v^n} Q^2}{m^2_{v^n}(Q^2 + m^2_{v^n})}\,, \ee
should be used instead to have a good convergence at low $Q^2$, where in the last line we have used
the sum rule (\ref{sumrule}).

Observe, however, that any truncation to a finite number of excited modes (as above) would eventually fail
for high $Q^2$, and summation over all modes would be required in order to achieve convergence.
There is an alternative way of doing this by noting that the form factor is proportional to the $Z$-average of
\be
\sum_{n\ge 1} {g_{v^n} \psi_{(2n-1)}(Z)\over Q^2+m^2_{v^{n}}}  \equiv G(Z,Q^2) \ .
\ee
Using the completeness relation, one can show that this function satisfies
\be
(1+Z^2)^{1\over 3}\partial_Z\left[(1+Z^2)\partial_Z G(Z,Q^2)\right] = \left({Q^2\over M_{KK}^2}\right) G(Z,Q^2)\quad,
\ee
with the boundary condition $G(Q^2,Z\to \pm \infty)=1$. In fact, this is nothing but the bulk-to-boundary propagator
(from the $Z\to \pm\infty$ boundary to the bulk $Z$ for the gauge field). It is not difficult to solve this equation numerically for each $Q^2$. We use this method in the numerical plots later.

Notice that the form factors $G_{2,3}(Q^2) $ are vanishing, since we are working at leading order in $N_c$, neglecting the subleading effects. In other words, $G_2(Q^2)$ and $G_3(Q^2)$ are expected to be of order $\cO(1/N_c)$ in the large $N_c$ limit. The fact that there is only one independent form factor was also observed in Ref.~\cite{Grigoryan:2007vg}, when discussing the form
factors of vector meson in the framework of AdS/QCD, and in Ref.~\cite{Hong:2007dq,Hashimoto:2008zw} for the nucleon form factors
in the Sakai-Sugimoto model.

The physically relevant magnetic dipole $G_{M}$, electric quadrupole $G_{E}$, and Coulomb quadrupole $G_{C}$
form factors, are predicted from holographic QCD to be:
\begin{align}\label{HoloMEQ}
G_{M}(Q^{2}) &=
\frac{m_N((3m_{\Delta} + m_N)(m_{\Delta} + m_N) + Q^{2})}{3m_{\Delta}(m_N + m_{\Delta})} \ G_{1}(Q^{2}) \ ,
\nonumber \\[5pt]
G_{E}(Q^{2}) &= \frac{m_N(m_{\Delta}^{2} - m_N^{2} - Q^{2})}{3m_{\Delta}(m_N + m_{\Delta})} \ G_{1}(Q^{2}) \ ,
\nonumber\\[5pt]
G_{C}(Q^{2}) &= \frac{4m_Nm_{\Delta}}{3(m_{\Delta} + m_N)} \ G_{1}(Q^{2}) \ .
\end{align}
As a result, the ratios take the following form:
\begin{align}\label{HoloRatios}
R_{EM}(Q^2) & = - \frac{(m_{\Delta}^{2} - m_N^{2} - Q^{2})}{(3m_{\Delta} + m_N)(m_{\Delta} + m_N) + Q^{2}}
\,,\\[5pt] \nonumber
R_{SM}(Q^2) & = -\sqrt{Q^2 +\frac{(m_{\Delta}^2- m_P^2-Q^2)^2}{4 m_{\Delta}^2}}
\frac{2m_{\Delta}}{(3m_{\Delta} + m_N)(m_{\Delta} + m_N) + Q^{2}} \ .
\end{align}
In case when $Q^2 = 0 $, we will have
\begin{align}\label{zeroHoloMEQ}
\frac{G_{M}(0)}{G_{1}(0)} = \frac{m_N(3m_{\Delta} + m_N)}{3m_{\Delta}} \ ,
%
%
\ \ \ \ \frac{G_{E}(0)}{G_{1}(0)} = \frac{m_N(m_{\Delta} - m_N)}{3m_{\Delta}} \ ,
%
%
\ \ \ \ \frac{G_{C}(0)}{G_{1}(0)} = \frac{4m_Nm_{\Delta}}{3(m_{\Delta} + m_N)} \ ,
\end{align}
\begin{align}\label{zeroHoloRatios}
R_{EM}(0) = R_{SM}(0) = - \frac{m_{\Delta} - m_N}{3m_{\Delta} + m_N} \ .
\end{align}
Recalling that baryon masses are of order $\cO(N_c)$, while $ \delta \equiv m_{\Delta} - m_N \sim \cO(1/N_c)$, we have
\begin{align}\label{degzeroHoloRatios}
R_{EM}(0) = R_{SM}(0) \simeq - \frac{\delta}{4m_N} \sim -\cO\left(\frac{1}{N_c^2}\right) \ .
\end{align}
Although, we work at leading order in $N_c$, this result is consistent, since holographic QCD predicts $G_{2,3} \sim \cO(1/N_c)$, and using Eqs.~(\ref{MEQ}) and (\ref{Ratios}), one can easily deduce Eq.~(\ref{degzeroHoloRatios}).
In agreement with our result, the $R_{EM}$
ratio for the $\gamma N \Delta$ transition was also shown to be of order $1/N_c^2$ in the Ref.~\cite{Jenkins:2002rj}.
Furthermore, the relation $R_{EM}(0) = R_{SM}(0)$ was also observed
in the Ref.~\cite{Pascalutsa:2007wz} within the large-$N_c$ limit (see also Ref.~\cite{Pascalutsa:2003zk}).
These observations provide an additional evidence that
the smallness of the $\gamma N \Delta$ $R_{EM}$ ratio is naturally explained in the large $N_c$ limit.

Finally, one can check that when $Q^2=0$,
\begin{align}
G_1(0) = \sqrt{2} \ \frac{\sqrt{30}}{4M_{KK}}\sum_{n \geq 1}  a_{\cV
v^n} \langle \psi_{(2n-1)}(Z)\rangle =  {\sqrt{15}\over 2} \ \frac{1}{M_{KK}} \ ,
\end{align}
where in the last step we used the sum rule from Eq.~(\ref{sumrule}). This result is universal for Type I, II
models. Therefore, from Eq.~(\ref{zeroHoloMEQ}) it follows that
\begin{align}
&G_{M}(0) = {\sqrt{15}\over 2} \ \frac{m_N}{M_{KK}} \frac{(3m_{\Delta} + m_N)}{3m_{\Delta}} \simeq 2.43\frac{m_N}{M_{KK}}  \ ,\nonumber\\
&G_{E}(0) = {\sqrt{15}\over 2}\ \frac{1}{M_{KK}} \frac{m_N(m_{\Delta} - m_N)}{3m_{\Delta}}  \simeq 0.154\frac{m_N}{M_{KK}} \ ,\nonumber \\
&G_{C}(0) = {\sqrt{15}\over 2} \ \frac{1}{M_{KK}}
\frac{4m_Nm_{\Delta}}{3(m_{\Delta} + m_N)} \simeq
1.47\frac{m_N}{M_{KK}} \ .
\end{align}
If we choose $M_{KK}=0.949$ GeV to fit the $\rho$-meson mass, and use the experimental nucleon and $\Delta$ baryon masses,
the above gives us
\begin{align}
G_{M}(0) \approx 2.41 \ ,\ G_{E}(0) \approx 0.153  \ , \ G_{C}(0)
\approx 1.46  \ .
\end{align}
However, since we are working in the large $N_c$ limit approximation, for consistency, the terms of order $ 1/N_c$ have to be dropped, and one should use the same mass for the nucleons and $\Delta$.
In this case, we will have $G_{E}(0) = 0$,
\begin{align}
G_{M}(0) = 2G_{C}(0) = \frac{4}{3}{\sqrt{15}\over 2} \
\frac{m_N}{M_{KK}} \ .
\end{align}
Numerically, we get $ G_M(0) = 2G_{C}(0) \simeq 2.58 $ and $ R_{EM}
= R_{SM} = 0 $.

\begin{figure}[t]
\begin{center}
\includegraphics[clip,width=12cm,angle=0]{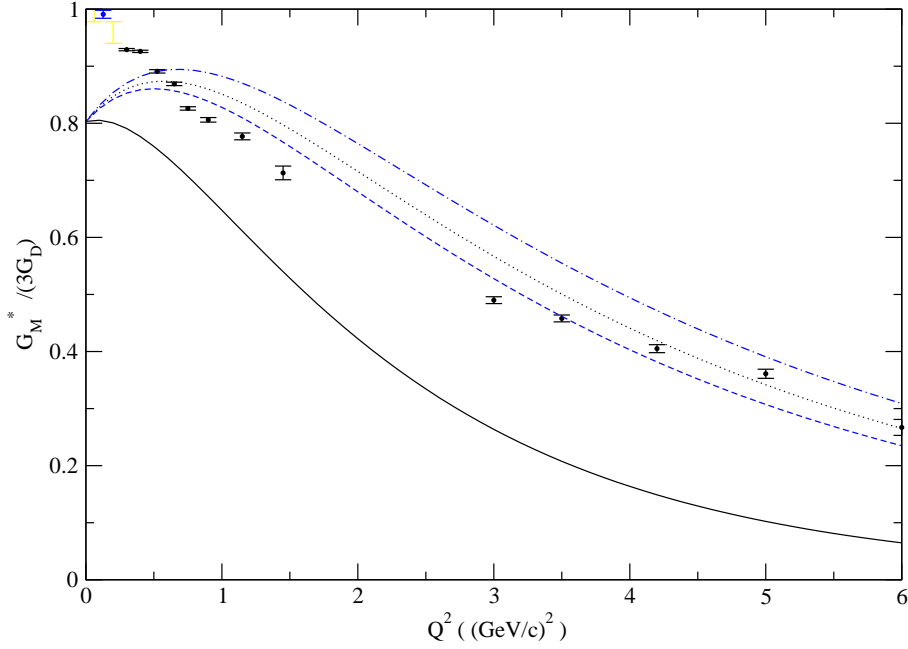}
\caption{The plot of the ratio $G_M^*(Q^2)/(3G_D(Q^2))$ as a function of $Q^2$, where $G_D(Q^2) = 1/(1 + Q^2/\Lambda^2)^2$ with
$\Lambda^2= 0.71 $ (GeV/c)$^2$.
The solid and dotted lines are the predictions from the holographic Type I and Type II models, respectively.
The dashed (dash-dotted) curves are from taking the parameter $a$ of Eq.~(29) to be
$20\%$ larger (smaller) than the value  $a \sim 0.240$.
The experimental data points are taken from \cite{clas-1}.} \label{fig1}
\end{center}
\end{figure}
We should point out that within the model one finds that the
quantized baryon mass is larger than the nucleon mass $m_N$. In the
5D effective field theory approach that is applied in
Refs.~\cite{Hong:2007kx,Hong:2007ay}, one finds
\be
{m_N\over M_{KK}} \approx 1.98 \ .
\ee
In order to obtain a better agreement in the baryon sector of the holographic model, different values for either $f_{\pi}$ or $M_{KK}$
have to be chosen. The ratio of the classical baryon mass $m_N \approx M_{B}^0 \sim \cO(N_c) $ to $M_{KK}$ scale (\ref{zeromass}),
before quantization is
\be
\frac{M_{B}^0}{M_{KK}} = 2\pi^3\left(\frac{f_{\pi}}{M_{KK}}\right)^2 \approx 0.59  \ .
\ee
This is clearly smaller than the ratio considered above, which may signal that the $1/N_c$ expansion and the numerical estimate for the baryon masses are no longer reliable. Since our predictions should be only leading order in $N_c$, there is no need to fit the parameters of the model ($f_{\pi}$ and $M_{KK}$) to the exact physical results. In particular, if we keep $M_{KK}=0.949$ GeV, while taking the baryon mass as input from experiment, this may correspond to changing the value for $f_{\pi}$, similar to the case in the Skyrme model \cite{Adkins:1983ya}. This issue is a problem of model and approximation that is being used. Whatever the resolution of this problem, it shouldn't affect our final results.

In Fig.1 we present a plot for the ratio $G_M^*(Q^2)/(3G_p^D(Q^2))$ that is commonly used in the literature, where $G_p^{D}(Q^2) = 1/(1 +
Q^2/\Lambda_p^2)^2 $ is the proton's empirical electric form factor with $ \Lambda^2_p \simeq 0.71 \ \GeV^2 $, and
\be
G_M^*(Q^2) = G_M(Q^2)/\sqrt{1+{Q^2\over (m_\Delta+m_N)^2 }} \ .
\ee

The data are taken from the experiments in \cite{clas-1}. The theoretical curves correspond to Type I and Type II models.
One may see that for $Q^2 \geq 0.5 \ \GeV^2 $ the better agreement with experiment is provided by the Type II model.
However, both models disagree with experiment (by about $20\%$) for $Q^2 \leq 0.5 \ \GeV^2$.
This suggests that, although smearing of baryons is required to get the correct behavior for $Q^2 \geq 0.5 \ \GeV^2 $, the exact account of $1/N_c$ corrections is required for lower energies.


\subsection{Helicity Amplitudes}

Equivalently, one can also parametrize the $\gamma^* N \Delta$ transition through the rest frame helicity amplitudes $A_{1/2}$ and
$A_{3/2}$  defined in terms of the following matrix elements of the electromagnetic current operator:
\begin{align}
A_{3/2} &\equiv -\frac{e}{\sqrt{2 q_{\Delta}}} \frac{1}{(4 M_N M_{\Delta})^{1/2}}\langle \Delta(\vec 0, +3/2)|{\bf J \cdot \epsilon}_{\lambda = +1} | N(-\vec q, \, +1/2)\rangle \ , \nonumber \\
A_{1/2} &\equiv -\frac{e}{\sqrt{2 q_{\Delta}}}\frac{1}{(4 M_N M_{\Delta})^{1/2}}\langle \Delta(\vec 0, +1/2)
|{\bf J \cdot \epsilon}_{\lambda = +1}| N(-\vec q, \, -1/2)\rangle \ ,
\label{eq:resthel}
\end{align}
where the transverse photon polarization vector entering in
$A_{1/2}$ and $A_{3/2}$ is given by ${\bf \epsilon}_{\lambda = +1} =
-1/\sqrt{2} (1, i, 0)$, the spin projections are along the $z$ axis
(along the virtual photon direction) and $q_\Delta$ is the magnitude
of the virtual photon three-momentum in the $
\Delta$ rest frame:
\begin{align}\label{defin}
q_{\Delta} \equiv |{\mathbf q}| = \frac{Q_+ Q_-}{2 M_\Delta} \ , \ \ \ Q_{\pm} \equiv \sqrt{(M_{\Delta} \pm M_N)^2 +Q^2} \ .
\end{align}
The helicity amplitudes are functions of the photon virtuality
$Q^2$, and can be expressed in terms of the Jones-Scadron
$\gamma^\ast N \Delta$ form factors as
\begin{align}\label{Helamp}
A_{3/2} &= - \cN \frac{\sqrt{3}}{2}\left(G_M + G_E\right) = - \cN \frac{\sqrt{3}}{2} G_M + \cO\left(\frac{1}{N^2_c}\right)  \ , \nonumber \\
A_{1/2} &= - \cN \frac{1}{2}\left( G_M - 3 G_E\right) =  - \cN \frac{1}{2}G_M + \cO\left(\frac{1}{N^2_c}\right) \ ,
\end{align}
where
\begin{align}
\cN \equiv \frac{e}{2} \left(\frac{Q_+ Q_-}{2 M_N^3}\right)^{1/2}\frac{(M_N + M_\Delta)}{Q_+} \ .
\end{align}
The above helicity amplitudes are expressed in units GeV$^{-1/2}$, and reduce at $Q^2 = 0$ to the photocouplings quoted by the
Particle Data Group~\cite{Amsler:2008zzb}. Experimentally, these helicity
amplitudes are extracted from the M1, E2, and C2 multipoles for the
$\gamma^* N \to \pi N $ process at the resonance position, i.e. for
$\pi N $ c.m. energy $W = M_{\Delta}$.

In terms of helicity amplitudes,
\begin{align}\label{ratio}
R_{EM} = \frac{A_{1/2} - \frac{1}{\sqrt{3}} \, A_{3/2}}{A_{1/2} + \sqrt{3}\, A_{3/2}} \ .
\end{align}
Notice that from the Eq.~(\ref{Helamp}) it follows that
\begin{align}
\frac{A_{3/2}}{A_{1/2}} \,=\, \sqrt{3} \,+\, {\mathcal O} \left(\frac{1}{N_c^2} \right) \ .
\end{align}
This result was also predicted in the Ref.~\cite{Jenkins:2002rj}
within the framework of the large $N_c$ QCD .

In the case $Q^2=0$, we have $Q_{\pm} = M_{\Delta} \pm M_N $, and
therefore
\begin{align}
\cN = \frac{e}{2} \left(\frac{M^2_{\Delta} - M^2_N}{2 M_N^3}\right)^{1/2} \simeq 0.094 \ \GeV^{-1/2} \ .
\end{align}
As a result, we will have from holographic QCD
\begin{align}
A_{1/2} &\simeq - 121 \ [10^{-3} \mathrm{GeV}^{-1/2}], \nonumber \\
A_{3/2} &\simeq - 209 \ [10^{-3} \mathrm{GeV}^{-1/2}], \nonumber \\
R_{EM} &\simeq  0 \ \%.
\label{pdg}
\end{align}
The experimental results (MAMI/A2 Collaboration \cite{Beck:1999ge} and LEGS Collaboration \cite{Blanpied:2001ae}),
taken from the Particle Data Group \cite{Amsler:2008zzb}, quote:
\begin{align}
A_{1/2} &= - \left( 135 \pm 6 \right)
\quad [10^{-3} \mathrm{GeV}^{-1/2}], \nonumber \\
A_{3/2} &= - \left( 250 \pm 8 \right)
\quad [10^{-3} \mathrm{GeV}^{-1/2}], \nonumber \\
R_{EM} &= - \left( 2.5 \pm 0.5 \right) \, \%.
\label{pdg}
\end{align}

From the values of the $\gamma^\ast N \Delta$ form factors at $Q^2 =
0$, one can extract some interesting static quantities. For the
dominant $M1$ transition, one can extract the static $N \to \Delta$
transition magnetic moment $\mu_{N \to \Delta}$ from the value of
$G_M(0)$ as~\cite{Tiator:2003xr}
\begin{align}
\mu_{N \to \Delta} = \sqrt{\frac{M_\Delta}{M_N}} \, G_M(0) \ ,
\end{align}
which is expressed in nuclear magnetons $\mu_N \equiv e / (2 M_N)$.
Furthermore, one can extract a static $N \to \Delta$ quadrupole
transition moment $Q_{N \to \Delta}$ as~\cite{Tiator:2003xr}
\begin{align}
Q_{N \to \Delta} = - 6 \sqrt{\frac{M_\Delta}{M_N}}\frac{1}{M_N q_\Delta(0)} G_E(0)  \ ,
\label{eq:qndel}
\end{align}
where $q_\Delta(0)$ is obtained from Eq.~(\ref{defin}) for $Q^2 = 0$,
as $q_\Delta(0) = (M_\Delta^2 - M_N^2)/ 2 M_\Delta$.

Our results from the holographic QCD framework, with masses of
baryons taken from experiments, are
\begin{align}
G_M(0) \simeq 2.41 \ , \ \ \ \mu_{N \to \Delta} \simeq 2.76 \mu_N \
, \ \ \ Q_{N \to \Delta} \simeq - 0.171 \ {\rm fm}^2  \ .
\end{align}
On the other hand, without taking experimental baryon
masses and neglecting terms of order $\cO(1/N_c)$, we will get
\begin{align}
G_M(0) \simeq 2.58 \ , \ \ \ \mu_{N \to \Delta} \simeq 2.58 \mu_N \
, \ \ \ Q_{N \to \Delta} \simeq 0 \ {\rm fm}^2 \ .
\end{align}
From the experiments, Ref.~\cite{Tiator:2000iy} extracted the values
\begin{align}
G_M(0) = \ 3.02 \pm 0.03 \ , \ \ \ \ \mu_{p \to \Delta^+} = \left[3.46 \pm 0.03 \right ] \mu_N \ , \ \ \
Q_{p \to \Delta^+} = - \left(0.0846 \pm 0.0033 \right) \ {\rm fm}^2 \ .
\end{align}
Taking into account that our results are of only leading order in
large $N_c$, we find about $20\%$ discrepancy with experiments as an
indication that the holographic model works consistently. It is an
important open problem to systematically improve the large $N_c$
expansion in the holographic QCD.

Transition amplitudes and their ratios were also discussed in the framework of the Skyrme like models, see e.g. Refs.~\cite{Wirzba:1986sc,Abada:1995db,Walliser:1996ps}.
In particular, Wirzba and Weise~\cite{Wirzba:1986sc} performed a modified Skyrme model calculation,
at leading order in $N_c$, where $R_{EM}$ takes values between $-2.5$\% and $-6$\%, depending on the coupling parameters
of the stabilizing terms.
In \cite{Walliser:1996ps}, Walliser and Holzwarth included rotational corrections, which are  of order $1/N_c$, and lead to a
quadrupole distortion of the classical soliton solution. Including such corrections, one finds a very good description of the
photocouplings and obtains a ratio $R_{EM} = -2.3 \%$, consistent with experiment.
Similarly, we also expect that quantum corrections, including rotational effects should improve our results and provide a better agreement with the experimental data.


\section{Conclusion}

Working in the framework of the holographic dual model of QCD proposed by Sakai and Sugimoto \cite{Sakai:2004cn,Sakai:2005yt}
with two massless flavors,
we consider the electromagnetic $N\to\Delta$ transition form factors at leading order in $N_c$.
By considering a relativistic generalization of the nonrelativistic vertices found in
Ref.~\cite{Park:2008sp} up to $1/N_c$ ambiguities, we treat the problem in a consistent relativistic way.
As a result of holographic computation, we establish that among three independent form factors, only one survives. Besides, the large $N_c$ dependence of transition form factors and their ratios coincide with what was expected in the earlier studies. In particular, the following fact was observed for the ratios: $R_{EM} = R_{SM} \sim \cO(1/N^2_c)$.

After employing the approximation where baryons are pointlike, we also consider a simple example, where the baryon wave functions are
smeared as a ground state oscillator {\it a la} Ref.~\cite{Hashimoto:2008zw}.
Although, the value of the $G_M(0)$ form factor remain the same for both models, we seem to get a better agreement
with experimental data for energies up to $6 \ \GeV^2$. This suggests that the finite size effects may indeed improve holographic
QCD predictions.
We leave the discussion of these effects for further studies.
Our most reliable results in this work are the values for the form factors obtained at $Q^2 = 0 $.

An interesting direction for further studies includes the possibility for studying transition form factors among various other excited hadron states. This approach can shed more light on photoproduction and electroproduction processes and help us to better understand the nature of baryon excited states.



\subsection*{Acknowledgments}
This work is supported partially by the U.S. Department of Energy, Office of Nuclear Physics Division, under
Contract No. DE-AC02-06CH11357.

Notes added.--Simultaneously with our work another article \cite{Ahn:2009px} appeared in the arXiv, discussing the same problem but in the framework of the holographic ``bottom-up'' model. Some of the main results as well as the hierarchies among the different form factors are qualitatively quite independent of the choice of the model.



\begin{thebibliography}{99}
\bibitem{Amsler:2008zzb}
  C.~Amsler {\it et al.}  [Particle Data Group],
  Phys.\ Lett.\  B {\bf 667}, 1 (2008).

\bibitem{Maldacena:1997re}
  J.~M.~Maldacena,
  Adv.\ Theor.\ Math.\ Phys.\  {\bf 2}, 231 (1998)
  [Int.\ J.\ Theor.\ Phys.\  {\bf 38}, 1113 (1999)];
  S.~S.~Gubser, I.~R.~Klebanov and A.~M.~Polyakov,
  Phys.\ Lett.\ B {\bf 428}, 105 (1998);
  E.~Witten,
  Adv.\ Theor.\ Math.\ Phys.\  {\bf 2}, 253 (1998).
%

\bibitem{Sakai:2004cn}
  T.~Sakai and S.~Sugimoto,
  Prog.\ Theor.\ Phys.\  {\bf 113}, 843 (2005).

%
\bibitem{Polchinski:2002jw}
  J.~Polchinski and M.~J.~Strassler,
  Phys.\ Rev.\ Lett.\  {\bf 88}, 031601 (2002);
  JHEP {\bf 0305}, 012 (2003);
%
  H.~Boschi-Filho and N.~R.~F.~Braga,
  JHEP {\bf 0305}, 009 (2003);
%
  J.~Erlich, E.~Katz, D.~T.~Son and M.~A.~Stephanov,
  Phys.\ Rev.\ Lett.\  {\bf 95}, 261602 (2005);
%
  L.~Da Rold and A.~Pomarol,
  Nucl.\ Phys.\ B {\bf 721}, 79 (2005);
%
  JHEP {\bf 0601}, 157 (2006);
%
  J.~Hirn and V.~Sanz,
  JHEP {\bf 0512}, 030 (2005).
%



\bibitem{Sakai:2005yt}
  T.~Sakai and S.~Sugimoto,
  Prog.\ Theor.\ Phys.\  {\bf 114}, 1083 (2005).


%
\bibitem{Brodsky:2003px}
  S.~J.~Brodsky and G.~F.~de T\'eramond,
  Phys.\ Lett.\ B {\bf 582}, 211 (2004);
%
  Phys.\ Rev.\ Lett.\  {\bf 94}, 201601 (2005);
%
  Phys.\ Rev.\ Lett.\  {\bf 96}, 201601 (2006);
%
  Phys.\ Rev.\  D {\bf 77}, 056007 (2008);
%
  Phys.\ Rev.\  D {\bf 78}, 025032 (2008);
%
  Phys.\ Rev.\ Lett.\  {\bf 102}, 081601 (2009).
%


\bibitem{Grigoryan:2007wn}
  H.~R.~Grigoryan and A.~V.~Radyushkin,
  Phys.\ Rev.\  D {\bf 76} (2007) 095007;
%
  Phys.\ Rev.\  D {\bf 76}, 115007 (2007);
%
  Phys.\ Rev.\  D {\bf 77}, 115024 (2008);
%
  Phys.\ Rev.\  D {\bf 78}, 115008 (2008);
%
  H.~R.~Grigoryan,
  Phys.\ Lett.\  B {\bf 662}, 158 (2008);
%
  Z.~Abidin and C.~E.~Carlson,
  Phys.\ Rev.\  D {\bf 77}, 095007 (2008);
%
  Phys.\ Rev.\  D {\bf 77}, 115021 (2008);
%
  H.~J.~Kwee and R.~F.~Lebed,
  JHEP {\bf 0801}, 027 (2008);
  Phys.\ Rev.\  D {\bf 77}, 115007 (2008).
%
\bibitem{Pomarol:2008aa}
  A.~Pomarol and A.~Wulzer,
  Nucl.\ Phys.\  B {\bf 809}, 347 (2009);
  G.~Panico and A.~Wulzer,
  arXiv:0811.2211 [hep-ph].

\bibitem{Hong:2006ta}
  D.~K.~Hong, T.~Inami and H.~U.~Yee,
  Phys.\ Lett.\  B {\bf 646}, 165 (2007).

\bibitem{Hong:2007dq}
  D.~K.~Hong, M.~Rho, H.~U.~Yee and P.~Yi,
  Phys.\ Rev.\  D {\bf 77}, 014030 (2008).

\bibitem{Hashimoto:2008zw}
  K.~Hashimoto, T.~Sakai and S.~Sugimoto,
  arXiv:0806.3122 [hep-th].
%
\bibitem{Kim:2008pw}
  K.~Y.~Kim and I.~Zahed,
  JHEP {\bf 0809}, 007 (2008).

\bibitem{Nawa:2006gv}
  K.~Nawa, H.~Suganuma and T.~Kojo,
  Phys.\ Rev.\  D {\bf 75}, 086003 (2007).

\bibitem{Hong:2007kx}
  D.~K.~Hong, M.~Rho, H.~U.~Yee and P.~Yi,
  Phys.\ Rev.\  D {\bf 76}, 061901 (2007).
%
\bibitem{Hata:2007mb}
  H.~Hata, T.~Sakai, S.~Sugimoto and S.~Yamato,
  arXiv:hep-th/0701280.

\bibitem{Hong:2007ay}
  D.~K.~Hong, M.~Rho, H.~U.~Yee and P.~Yi,
  JHEP {\bf 0709}, 063 (2007).

\bibitem{Park:2008sp}
  J.~Park and P.~Yi,
  JHEP {\bf 0806}, 011 (2008).
%
\bibitem{Hata:2008xc}
  H.~Hata, M.~Murata and S.~Yamato,
  arXiv:0803.0180 [hep-th].


\bibitem{Seki:2008mu}
  S.~Seki and J.~Sonnenschein,
  arXiv:0810.1633 [hep-th].

\bibitem{Kim:2008iy}
  K.~Y.~Kim and I.~Zahed,
  arXiv:0901.0012 [hep-th];
  K.~Hashimoto, T.~Sakai and S.~Sugimoto,
  arXiv:0901.4449 [hep-th];
  Y.~Kim, S.~Lee and P.~Yi,
  arXiv:0902.4048 [hep-th].

\bibitem{Carlson:1985mm}
  C.~E.~Carlson,
  Phys.\ Rev.\ D {\bf 34}, 2704 (1986);
%
  C.~E.~Carlson and J.~L.~Poor,
  Phys.\ Rev.\ D {\bf 38}, 2758 (1988).

\bibitem{Beck:1997ew}
  R.~Beck {\it et al.},
  Phys.\ Rev.\ Lett.\  {\bf 78}, 606 (1997).

\bibitem{Becchi:1965}
C. Becchi and G. Morpurgo,
 Phys.\ Lett. {\bf 17}, 352 (1965).

\bibitem{Aznaurian:1993rk}
  I.~G.~Aznaurian,
  Phys.\ Lett.\ B {\bf 316}, 391 (1993).


\bibitem{Jenkins:2002rj}
  E.~E.~Jenkins, X.~d.~Ji and A.~V.~Manohar,
  Phys.\ Rev.\ Lett.\  {\bf 89}, 242001 (2002).



\bibitem{Belyaev:1995ya}
  V.~M.~Belyaev and A.~V.~Radyushkin,
  Phys.\ Rev.\ D {\bf 53}, 6509 (1996).

\bibitem{Braun:2005be}
  V.~M.~Braun, A.~Lenz, G.~Peters and A.~V.~Radyushkin,
  Phys.\ Rev.\  D {\bf 73}, 034020 (2006).

\bibitem{Wang:2009ru}
  L.~Wang and F.~X.~Lee,
  Phys.\ Rev.\  D {\bf 80}, 034003 (2009).

\bibitem{Alexandrou:2004xn}
C.~Alexandrou, P.~de Forcrand, H.~Neff, J.~W.~Negele, W.~Schroers and A.~Tsapalis,
Phys.\ Rev.\ Lett.\  {\bf 94}, 021601 (2005);
%
  C.~Alexandrou, G.~Koutsou, H.~Neff, J.~W.~Negele, W.~Schroers and A.~Tsapalis,
Phys.\ Rev.\  D {\bf 77}, 085012 (2008).


\bibitem{Review}
D. Drechsel and L. Tiator,
J. Phys. G {\bf 18}, 449 (1992);
%
  B.~Krusche and S.~Schadmand,
  Prog.\ Part.\ Nucl.\ Phys.\  {\bf 51}, 399 (2003);
%
  V.~D.~Burkert and T.~S.~H.~Lee,
  Int.\ J.\ Mod.\ Phys.\ E {\bf 13}, 1035 (2004);
%
  V.~Pascalutsa, M.~Vanderhaeghen and S.~N.~Yang,
  Phys.\ Rept.\  {\bf 437}, 125 (2007).
%
\bibitem{Son:2003et}
  D.~T.~Son and M.~A.~Stephanov,
  Phys.\ Rev.\  D {\bf 69}, 065020 (2004).

\bibitem{Adkins:1983ya}
  G.~S.~Adkins, C.~R.~Nappi and E.~Witten,
  Nucl.\ Phys.\  B {\bf 228}, 552 (1983).

\bibitem{Amado:1986ef}
  R.~D.~Amado, R.~Bijker and M.~Oka,
  Phys.\ Rev.\ Lett.\  {\bf 58}, 654 (1987).

\bibitem{Gazit:2008gz}
  D.~Gazit and H.~U.~Yee,
  Phys.\ Lett.\  B {\bf 670}, 154 (2008).



\bibitem{Rarita:1941mf}
  W.~Rarita and J.~S.~Schwinger,
  Phys.\ Rev.\  {\bf 60}, 61 (1941).


\bibitem{Johnson:1960vt}
  K.~Johnson and E.~C.~G.~Sudarshan,
  %
  Annals Phys.\  {\bf 13}, 126 (1961).

\bibitem{Hagen:1972ea}
  C.~R.~Hagen,
  Phys.\ Rev.\ D {\bf 4}, 2204 (1971).

\bibitem{Velo:1969bt}
  G.~Velo and D.~Zwanziger,
  Phys.\ Rev.\  {\bf 186}, 1337 (1969).

\bibitem{Singh:1973gq}
  L.~P.~S.~Singh,
  Phys.\ Rev.\ D {\bf 7}, 1256 (1973).


\bibitem{Hagen:1982ez}
  C.~R.~Hagen and L.~P.~S.~Singh,
  Phys.\ Rev.\ D {\bf 26}, 393 (1982);
V.~Pascalutsa,
Phys.\ Rev.\ D {\bf 58}, 096002 (1998);
%
  S.~Deser, V.~Pascalutsa and A.~Waldron,
  Phys.\ Rev.\ D {\bf 62}, 105031 (2000);
%
  T.~Pilling,
    Mod.\ Phys.\ Lett.\ A {\bf 19}, 1781 (2004);
  Int.\ J.\ Mod.\ Phys.\ A {\bf 20}, 2715 (2005);
%
  N.~Wies, J.~Gegelia and S.~Scherer,
  Phys.\ Rev.\ D {\bf 73}, 094012 (2006);
%
V.~Pascalutsa and R.~G.~E.~Timmermans,
Phys.\ Rev.\ C {\bf 60}, 042201(R) (1999).


\bibitem{Jones:1972ky}
  H.~F.~Jones and M.~D.~Scadron,
  Annals Phys.\  {\bf 81}, 1 (1973).

\bibitem{Buchmann:2004ia}
  A.~J.~Buchmann,
  Phys.\ Rev.\ Lett.\  {\bf 93}, 212301 (2004).

\bibitem{Caia:2004pm}
  G.~L.~Caia, V.~Pascalutsa, J.~A.~Tjon and L.~E.~Wright,
  Phys.\ Rev.\ C {\bf 70}, 032201 (2004).


\bibitem{Pascalutsa:2005ts}
  V.~Pascalutsa and M.~Vanderhaeghen,
  Phys.\ Rev.\ Lett.\  {\bf 95}, 232001 (2005).


%
\bibitem{Grigoryan:2007vg}
  H.~R.~Grigoryan and A.~V.~Radyushkin,
  Phys.\ Lett.\  B {\bf 650}, 421 (2007).
%

\bibitem{Pascalutsa:2007wz}
  V.~Pascalutsa and M.~Vanderhaeghen,
  Phys.\ Rev.\  D {\bf 76}, 111501 (2007).

\bibitem{Pascalutsa:2003zk}
  V.~Pascalutsa and D.~R.~Phillips,
  Phys.\ Rev.\ C {\bf 68}, 055205 (2003).

\bibitem{clas-1}
K. Joo et al. CLAS Collaboration, Phys. Rev. Lett. {\bf 88}, 122001 (2003);
Phys. Rev. {\bf C68}, 032201 (2003);
%
M. Ungaro et al. CLAS Collaboration,
Phys. Rev. Lett. {\bf 97}, 112003 (2006);
%
L.C. Smith et al. CLAS Collaboration, Proceeding of Workshop
``Shape of Nucleons'', Athen, (2006);
%
N.F. Sparveris et al.,  Phys. Rev. Lett. {\bf 94}, 022003 (2005);
%
S. Stave et al. Eur. Phys. J. A. {\bf 30}, 471 (2006);
N.F. Sparveris et al., Phys. Lett {\bf B651}, 102 (2007).


\bibitem{Beck:1999ge}
  R.~Beck {\it et al.},
  Phys.\ Rev.\ C {\bf 61}, 035204 (2000).

\bibitem{Blanpied:2001ae}
  G.~Blanpied {\it et al.},
  Phys.\ Rev.\ C {\bf 64}, 025203 (2001).


 \bibitem{Tiator:2003xr}
  L.~Tiator, D.~Drechsel, S.~S.~Kamalov and S.~N.~Yang,
  Eur.\ Phys.\ J.\ A {\bf 17}, 357 (2003).

\bibitem{Tiator:2000iy}
  L.~Tiator, D.~Drechsel, O.~Hanstein, S.~S.~Kamalov and S.~N.~Yang,
  Nucl.\ Phys.\ A {\bf 689}, 205 (2001).

\bibitem{Wirzba:1986sc}
  A.~Wirzba and W.~Weise,
  Phys.\ Lett.\ B {\bf 188}, 6 (1987).

\bibitem{Abada:1995db}
  A.~Abada, H.~Weigel and H.~Reinhardt,
  Phys.\ Lett.\ B {\bf 366}, 26 (1996).

\bibitem{Walliser:1996ps}
  H.~Walliser and G.~Holzwarth,
  Z.\ Phys.\ A {\bf 357}, 317 (1997).

\bibitem{Ahn:2009px}
  H.~C.~Ahn, D.~K.~Hong, C.~Park and S.~Siwach,
  arXiv:0904.3731 [hep-ph].



\end{thebibliography}
\end{document}